\documentclass[]{article}

\pdfoutput=1
\usepackage{amssymb,amsmath}
\usepackage{graphicx}
\usepackage{natbib}
\usepackage[usenames,dvipsnames]{color}
\usepackage[notref,notcite,final]{showkeys}
\usepackage[pagewise,displaymath,modulo]{lineno}
\newcommand{\degc}{^\circ\text{C}}
\newcommand{\reffig}[1]{Fig.~\ref{#1}}
\newcommand{\refeq}[1]{Eq.~\ref{#1}}
\newcommand{\refsec}[1]{Sec.~\ref{#1}}

\author{Stefan Siegert\footnote{College of Engineering, Mathematics and Physical Sciences, University of Exeter, Harrison Building, North Park Road, Exeter, EX4 4QF, United Kingdom, Email: s.siegert@exeter.ac.uk}, Jochen Br\"ocker\footnote{University of Reading, Whiteknights, PO Box 220, Reading, RG6 6AX, United Kingdom, Email: j.broecker@reading.ac.uk}, Holger Kantz\footnote{Max Planck Institute for the Physics of Complex Systems, N\"othnitzer Str. 38, 01189 Dresden, Germany, Email: kantz@pks.mpg.de}}

\title{Skill of data based predictions versus dynamical models --
case study on extreme temperature anomalies}

\begin{document}

\maketitle

\begin{abstract}
We compare probabilistic predictions of extreme temperature anomalies issued by
two different forecast schemes. One is a dynamical physical weather model, the
other a simple data model. We recall the concept of skill scores in order to
assess the performance of these two different predictors.  Although the result
confirms the expectation that the (computationally expensive) weather model
outperforms the simple data model, the performance of the latter is surprisingly
good. More specifically, for some parameter range, it is even better than the
uncalibrated weather model. Since probabilistic predictions are not easily
interpreted by the end user, we convert them into deterministic yes/no
statements and measure the performance of these by ROC statistics. Scored in
this way, conclusions about model performance partly change, which illustrates
that predictive power depends on how it is quantified.
\end{abstract}

\section{Introduction}

In this contribution, as in most others of this collection of articles, Extreme
Events are short-lived large deviations from a system's normal state. More
precisely, at least one relevant system variable or an order parameter (the
latter being synonymous with ``observation'', which is a way to characterise a
microstate of a system on macroscopic scales) assumes a numerical value which
is either much bigger or much smaller than ``on average''. Without being more
specific, one might assume that such a value occurs in the tail of the
probability distribution for this quantity, and that ``extreme'' means to
observe a deviation from the mean which exceeds typical deviations. Hence,
Extreme Events are inevitably also rare events. 

For some phenomena, there are active debates on whether or not extremes occur
more frequently than in a Gaussian distribution (e.~g., for rogue waves
\citep{roguewaves}). Indeed, for distributions with fat tails such as
L\'evy-stable distributions with $\alpha < 2$, power law tails lead to a much
larger number of extremes and to a considerable percentage of extremes which
are by orders of magnitude larger than normal events, as compared to Gaussian
distributions. Since such distributions have diverging higher moments, this
situation can robustly be detected in time series data by a lack of convergence
of finite-time estimates of those moments. This can be nicely illustrated in
earthquake time series, when released energy is considered instead of
magnitudes. The running mean of the energy per event increases with every major
earthquake and therefore does not converge to a finite value. This is in
agreement with the fact that the probability distribution for released energy
is observed to be a power law, $p(E) \propto E^{-\beta}$, with $\beta\approx
0.5$, which does not have a finite mean. 

With the (trivial) observation that distributions and in particular the
existence of higher moments is not a property which is invariant under
nonlinear transformations of variables, it is not surprising that there are
many natural phenomena where empirical magnitude-frequency distributions
suggest that the underlying true distribution does not have diverging higher
moments. We found that wind gusts even show exponential distributions,
precipitation data do have some outliers in an otherwise exponential
distribution, which cannot be easily interpreted, and river levels have a
finite maximum. Nonetheless, also phenomena such as wind gusts, precipitation,
air pressure and other atmospheric data can be studied under the aspect of
extreme events.  Moreover, passing over to a different macroscopic quantity,
e.~g., the induced costs due to damage associated with a natural extreme event,
evidently changes the nature of a magnitude distribution.

In summary, in this contribution we will consider events as being extreme,
whenever they are in the uppermost or lowermost range of values for a given
quantity, regardless of how large the deviation from the mean value is. More
specifically, we will discuss below extreme temperature anomalies, i.~e., large
deviations of the surface temperature from its climatological average for the
corresponding day of the year, which are to a good approximation Gaussian
distributed. We consider the performance of predictors for the temperature
anomaly to overcome a given threshold on the following day for all possible
threshold values. Under this setting one can speak of ``prediction of extreme
events'' only in the limit of this threshold being very high, or, respectively,
in the limit that the average event rate goes to zero. The unexpected result of
this case study will be that the performance in this limit will differ when
being measured through different scoring schemes, and that it is therefore not
evident how predictable such extremes really are in an abstract, non-technical
sense (for every precisely defined scoring scheme, there is certainly a precise
number which characterises predictability). The other issue of this article
will be to compare sophisticated physical dynamical models to simple data based
predictors. Here, the conclusion is that physical dynamical models are usually
better than data based predictions. However, there are exceptions, and we
present examples where a simple data-driven model outperforms a
physical-dynamical weather model.

\section{Forecast concepts}

Prediction implies that we issue some statement about the future, based on
information from the past, and that there is a time interval between the time
when issuing the prediction and the time for which the prediction should apply.
In weather forecasting this is called the lead time, in other contexts it is
called prediction horizon. The measurement against which the prediction is
eventually compared is called the verification.  Prediction targets can be
either discrete or continuous. In the former case, the target variable can take
on only a finite number of values. In the case of only two possibilities, we
speak of a binary target. A continuous target can take on an infinite number of
possible values. In the context of weather forecasting, the target event
``above or below $30\degc$'' is binary, ``cold/mild/warm/hot'' (defined by
precise temperature ranges) is discrete, and predicting an exact temperature
value is a continuous prediction target. For each of these targets, predictions
can be either deterministic or probabilistic. Deterministic prediction involves
a dogmatic statement about the target event, such as ``it will be above
$30\degc$ tomorrow at noon'' (binary), or ``the temperature in two days at noon
will be exactly $35\degc$'' (continuous). Probabilistic predictions, on the
other hand, assign probabilities to express degrees of (un-)certainty about the
prediction target.  A binary probabilistic prediction is for example ``the
probability of having above $30\degc$ tomorrow at noon is 70\%'', and a
continuous probabilistic prediction is ``the probability distribution $p(T)$
assigned to tomorrow's temperature $T$ at noon is a Gaussian with parameters
$\mu=32\degc$ and $\sigma=2\degc$''. Furthermore, prediction targets can refer
to a given moment in time, or to a time interval, to a fixed location or to a
geographical region, etc. Another extension is to consider multivariate
variables such as wind velocity vectors or temperature fields. The actual
realization of the target variable, the measurement against which the
prediction is eventually compared, is referred to as the verification. The
above discussion highlights that in every prediction problem a precise
definition of the prediction target is crucial and not completely trivial, a
point which might not be obvious at first sight.

Every forecasting algorithm is an input - output relation, where inputs are
variables which characterise the knowledge about the system under concern at
time $t$, and the output is one of the forecast products discussed above.
Already for a given set of input data and the same prediction target, one can
design very different ways to actually produce a specific value for the output.
The simplest forecast is a constant value independent of any inputs.
This can make sense, e.g., in the case of continuous deterministic forecasts
and of probabilistic binary forecasts. For the deterministic forecast, it could
be the mean value of the prediction target (or should it be its median?), and
for the probabilistic forecast it could be the average frequency of occurrence
of the target. But notice that already for this very simple scheme the
optimality of a specific value depends on the way how the performance of a
forecast is measured (e.~g., whether to use the mean or the median depends on
the performance measure). As a further complication, different forecast schemes
for the very same target might use different sets of input variables.

There are many methods to detect and describe dependencies between input data
and the target value on a training set. These include time series models,
regression models, decision trees, or neural networks, just to name a few. In
climate research, where physical models of the atmosphere-ocean systems are
employed, the models differ in the way how different physical processes are
resolved and how the non-resolved processes are parametrized, but also in the
spatial and temporal resolution of the models. 

In this contribution we will focus on two types of predictions, which we
evaluate by two different types of performance measures. One prediction will be
a probabilistic forecast for a binary event, which issues a probability $p$ for
``yes'' and accordingly $1-p$ for ``no''. The other will be a binary
deterministic prediction which will either predict ``yes'' or ``no'', and it will be
derived from the probabilistic forecast. These two types of predictions will be
evaluated by proper skill scores and ROC (Receiver Operating Characteristic) analysis, respectively.

Our target is the prediction of weather extremes. Since true weather extremes
are rare and any statistical analysis of the performance of any predictor is
therefore strongly error prone, we will relax the requirement of ``extreme'' a
bit and at the same time we will look at a quantity which exhibits ``extremes''
independent of season: We will study the fluctuations of temperature anomalies,
and the prediction target is that the anomaly will exceed a fixed threshold
on the next day, given that the anomaly of the present day is below that
threshold. The latter restriction - prediction only if current temperature
anomaly is ``not extreme'' - takes into account the aspect of ``event'': Even
though a heat wave, say, typically lasts several days, prediction of its onset
seems to be more interesting than the prediction that it will continue on the
next day. As said, we concentrate on temperature anomalies, which are the
differences of the actual temperature and the climatological mean temperature
at the given day of the year. Therefore, an extreme anomaly can occur at any
season and hence the event rate is independent of the current season. 

We will use two types of models for performing predictions: Simple data models,
where we predict the temperature anomaly of the next day based only on
measurement data, with an interdependence structure which is extracted from a
long dataset of historic recordings. The other model type relies on a global
general circulation model, i.~e., a weather model which is fed with station
data from the entire globe, and which contains a good portion of the physics of
the atmosphere.

With these two types of model, we will predict the probability that the
temperature anomaly will exceed a given threshold 24h ahead, if it is below
that threshold at the time when the forecast is issued. Later we will convert
these predicted probabilities into binary deterministic forecasts. Given the
fact that the weather model is by many orders of magnitude more complex and
more costly than the data model, and that it also contains a factor of (at
least) 10$^5$ more input data which characterize the current state of the
atmosphere, we expect that it will outperform the data model by orders of
magnitude, but by how many orders? This case study will give some surprising
results.

\section{The data}\label{sec01}

We consider a data set of temperature observations at 2 meters height for the
location Hannover, Germany. The data set is provided by the DWD climatological data base \citep{DWDclim}.
It consists of $N=23741$ daily temperature measurements $T'_n$ taken at 12:00
UTC between 1946 and 2010. The time index $n$ thus indicates ``days since
1946/01/01''. The mean and variance of the time series are $\overline{T'} =
12.06\degc$ and $\overline{(T'-\overline{T'})^2} = (8.31\degc)^2$,
respectively.

\begin{figure}
\centering
\includegraphics[width=0.7\columnwidth]{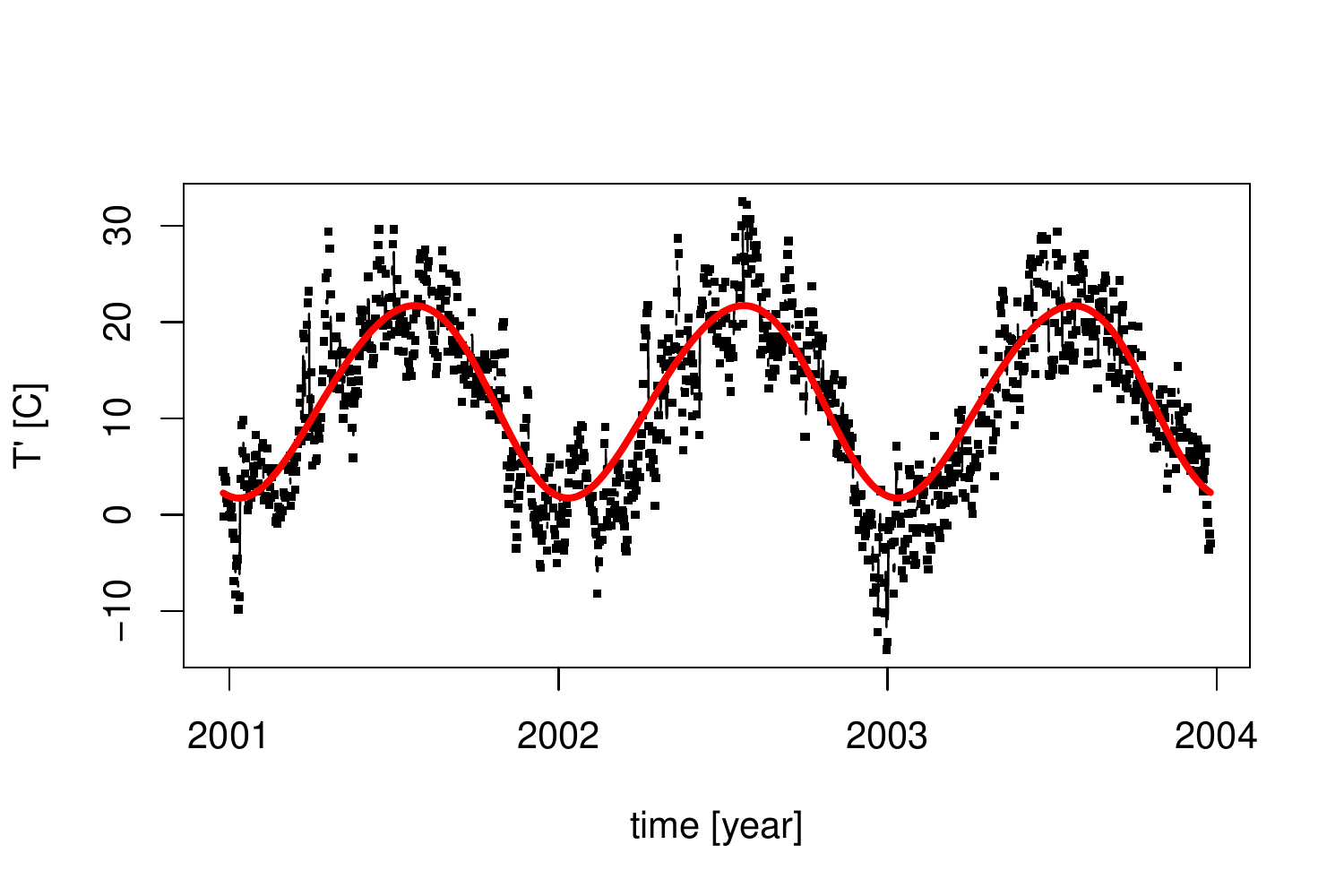}
\caption{Black markers show a three-year sample of the daily temperature data
for Hannover (Germany), which we analyze in this study. The fitted climatology
$c_n$ is shown as a red line.}
\label{fig01}
\end{figure}

Since the number of really extreme surface temperature events, i.~e., the
number of exceptionally cold or exceptionally hot days per year, is rather
small and clearly restricted to summer, resp. winter season, we consider
\textit{anomalies}. The anomalies $T_n$ are defined as the deviation of the
actual temperature $T'_n$ from a typical, expected temperature value $c_n$,
called the \textit{climatology}. A pragmatic approach to estimate the
climatology for day $n$ is to average the observed temperature values on the
same date over a number of previous years. However, the result even for 64
years (as they are available to us) is not a smooth function of $n$, as one
would assume.  We implement this smoothness assumption by modeling the
climatology as a seasonal cycle which is composed of a constant component, a
component proportional to $\sin(\omega n +\phi_1)$, and a component
proportional to $\sin(2\omega n +\phi_2)$, where $\omega =2\pi/(365.2425 \mbox{
days})$ is the rotational frequency of the Earth and $\phi_1$ and $\phi_2$ are
phases that have to be estimated along with the proportionality constants.
Higher harmonics could be taken into account as well but here we restrict the
estimator to only the first two. The seasonal cycle $c_n$ is estimated by
choosing a coefficient vector $\boldsymbol\beta=(\beta_0,\cdots,\beta_4)$ such
that the sum of squared differences between the observed temperatures $T'_n$
and
\begin{equation}
c_n = \beta_0 + \beta_1 \cos{\omega n} + \beta_2 \sin{\omega n} + \beta_3
\cos{2\omega n} + \beta_4 \sin{2 \omega n}
\end{equation}
is minimized. For the Hannover temperature time series (using $n=1, \cdots, N$)
the least squares fit of $\boldsymbol\beta$ is given by
$\hat{\boldsymbol\beta}=(12.1,-9.5,-2.9,-0.6,0.2)$. The temperature anomalies
are then constructed from the observed data and the climatology by
\begin{equation}
T_n=T'_n-c_n.
\end{equation}
A three-year sample of the temperature data and the fitted seasonal cycle are
shown in \reffig{fig01}. The anomaly $T_n$ is what we consider the non-trivial
part of the temperature, the part that can not be easily predicted since it is
strongly fluctuating.  Our goal will be to predict whether or not future
anomalies exceed some (possibly high) threshold, given that the current anomaly
is below that threshold.

\begin{figure}
\centering
\includegraphics[width=0.7\columnwidth]{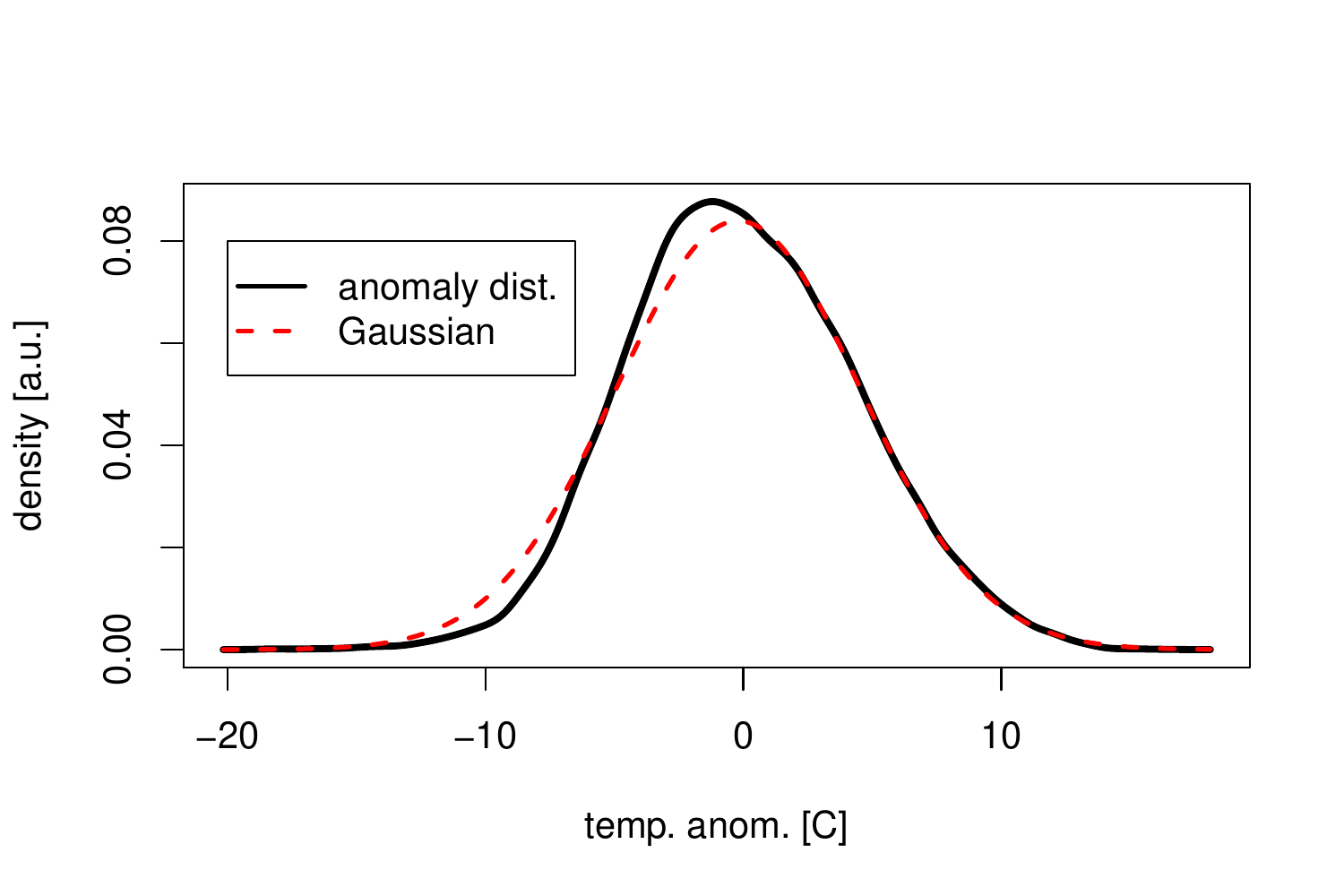}
\caption{Temperature anomalies are calculated by subtracting the climatology
from the temperature time series. The anomaly distribution reconstructed by
kernel density estimation (black line) is approximately Gaussian (red line).
Negative anomalies appear more concentrated towards zero than what would be
expected in a Gaussian distribution.}
\label{fig02}
\end{figure}

The distribution of the temperature anomalies $T_n$ is approximately Gaussian,
as shown in \reffig{fig02}. The density was fit using Gaussian kernel density
estimation with automatic bandwidth selection, as implemented by the {\tt
R}-function {\tt density} provided by the {\tt stats}-package (\cite{Rmanual},
see also \cite{silverman1998density}).  The distribution differs from a
Gaussian in that it is slightly right-skewed, indicating that the negative
anomalies are less variable than the positive ones. In a log-normal plot (not
shown), the tails of the fitted distribution decay even faster than that of the
Gaussian. This is an artifact of the density estimation procedure where the
tail behavior of the reconstruced distribution is governed by the tail of the
kernel, which has a much smaller variance than the data whose distribution is
estimated. We can thus not draw definite conclusions about the true tail
behavior of our data.

\begin{figure}
\centering
\includegraphics[width=0.7\columnwidth]{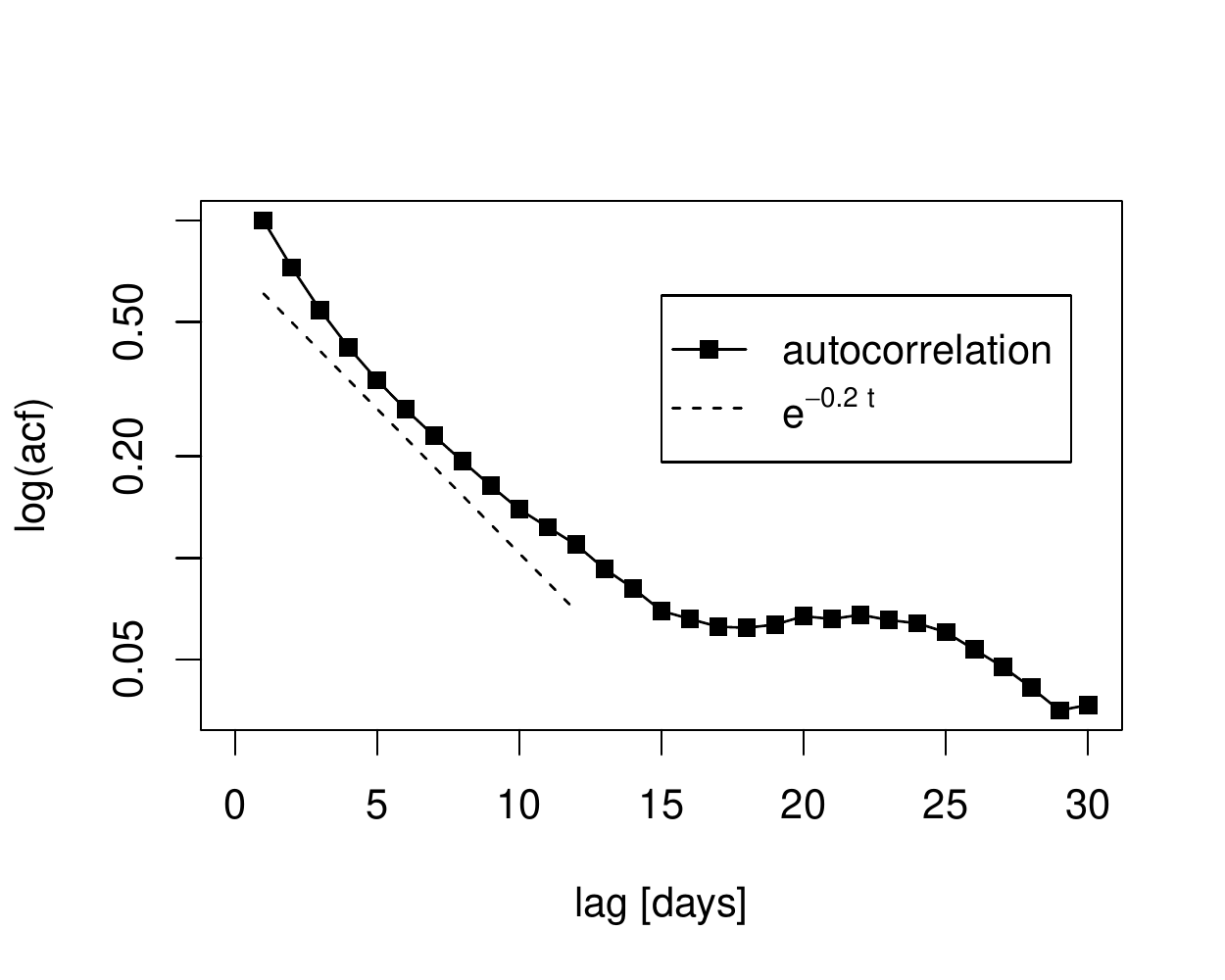}
\caption{Autocorrelation function of the temperature anomalies $T_n$ 
plotted in log-normal axes. It 
decays exponentially with a decay time of about 5 days. }
\label{fig03}
\end{figure}

The autocorrelation function of the temperature anomaly exhibits an
approximately exponential decay with a decay time of about 5 days as shown in
\reffig{fig03}. The non-vanishing autocorrelation function for small lags
indicates that the value of the anomaly at time $n$ contains predictive
information about the value of the anomaly at time $n+1$. So, evidently,
temperature anomalies are not white noise.

\section{The forecast models}

In order to make forecasts about the future, we need models of how information
about the future is computed from knowledge about the present situation. In the
following, we start from the simplest one-parameter model, then introduce a
dynamical data based model, and a complex weather model, together with an
additional adjustment to observed data, so that we have a total of 4 models to
be compared.

\subsection{The base rate model\label{sec:CEBR}}

A data model extracts details of the dependencies between successive values of
a time series which we can use for prediction. Many different such models
co-exist. In machine learning, rather general but parameterised input-output
relations are used. Learn pairs of input and corresponding output are used to
adapt the model parameters to the observed data. Alternatively, one can use
well established dynamical models, where the class of linear Gaussian models is
the most prominent. In fact, in the following subsection, we will argue that a
simple AR(1) process is an excellent compromise between model complexity and
accuracy. Here, we start with an even simpler model. 

A base rate model relies on the (known) average event rate $r$. It issues
predictions which are independent of time and independent of the present state
of the true system, simply predicting that the event will take place with
probability $r$.

We want to predict whether the anomaly will exceed a certain threshold on the
next day. But we are only interested in such a prediction if the present
anomaly is below that threshold. This latter complication takes into account
that we are interested in the prediction of ``events'', i.~e., of something
that is a change with respect to the current situation. Therefore, we will not
make any prediction at times $n$ at which the anomaly is already above
threshold, which means that for low thresholds we will have a strongly reduced
number of prediction trials. The distribution of the temperature anomalies only over the days on which a forecast is issued has a cut-off at the value of the threshold. Due to autocorrelation, the anomaly distribution over days for which a prediction was issued has a smaller mean than the unconditional anomaly distribution shown in \reffig{fig02}.

In this binary prediction, an event $X_{n+1}=1$ is observed, whenever
$T_{n+1}>\tau$ for a threshold value $\tau$, but we make a prediction only if
$T_{n}\le\tau$, that is, if $X_n=0$. Therefore, the event rate evaluated on
$N$ data is given by
\begin{equation}
r_\tau = \mathbb{P}(T_{n+1}>\tau \mid T_n\le\tau)\approx \frac{\sum_{n=1}^{N-1} (1-X_n)X_{n+1}}{\sum_{n=1}^{N-1}(1-X_n)}\;.\label{eq:empiricalrate}
\end{equation}
This base rate model will now predict that, given that $X_n=0$ (i.~e.,
$T_n\le\tau$), then the anomaly on the following day will exceed the threshold
$\tau$ with a probability $r_\tau$, independent of any information about the
current weather. If $X_n=1$ ($T_n>\tau$), no prediction will be made and the
corresponding day is not considered a forecast instance. Therefore, we will refer
to this model as the {\sl conditional exceedance base rate} (CEBR) model.

The base rate model is the simplest model one can think of and it will
therefore serve as a benchmark. The only parameter of this model is the rate
$r_\tau$, which can be easily extracted from recorded data. In this sense, it
is a purely data-driven model. Let us stress that a sophisticated weather model
as it will be described below creates weather predictions through modeling of
physical processes, and that there is no guarantee that such a model generates
events with the correct base rate. Therefore, the benchmark provided by the
base rate model is a serious one.

\subsection{The AR(1) model}

A more reasonable model than the base rate model should take into account our
knowledge about the current weather state and thereby yield predictions which
vary along the time axis. Based on the almost-Gaussianity of the temperature
anomalies, and based on their almost-exponentially fast decay of
autocorrelations, a reasonable model which makes use of current and past
observations is a linear autoregressive (AR) process:
\begin{equation}
T_{n+1} = \mu + \sum_{i=1}^p \alpha_i T_{n+1-i} + \epsilon_n,
\end{equation}
where $p$ is the order of the model, $\alpha_i$ are the constant AR parameters
and the residuals $\epsilon_n$ are white noise with zero mean and variance
$\sigma^2$. We assume for the mean $\mu=0$, because we consider anomalies
whose mean is zero by construction. Given the order $p$, the parameters could
be adapted by minimisation of the root mean squared prediction error with
respect to the $\alpha_i$, or some modifications such as the Yule Walker equations.
Different sophistications for the estimate of AR-coefficients and also of the
order of AR-models exist \citep{schelter2007handbook}. 

We split the full data set into a training and a test set, i.~e. we fit the
model coefficients on data from 1946 to 1978 (inclusive), and make predictions
and compute their performance on the remaining data from 1979 onward. We use
the {\tt R}-function {\tt ar} provided by the {\tt stats}-package
\citep{Rmanual} to fit the AR parameters $\alpha_i$ as well as the variance of
the residuals $\sigma^2$ using maximum likelihood estimation. An optimal order
of $p=6$ is suggested by Akaike's information criterion \citep{akaike1974new}. 

\begin{figure}
\centering
\includegraphics[width=0.7\columnwidth]{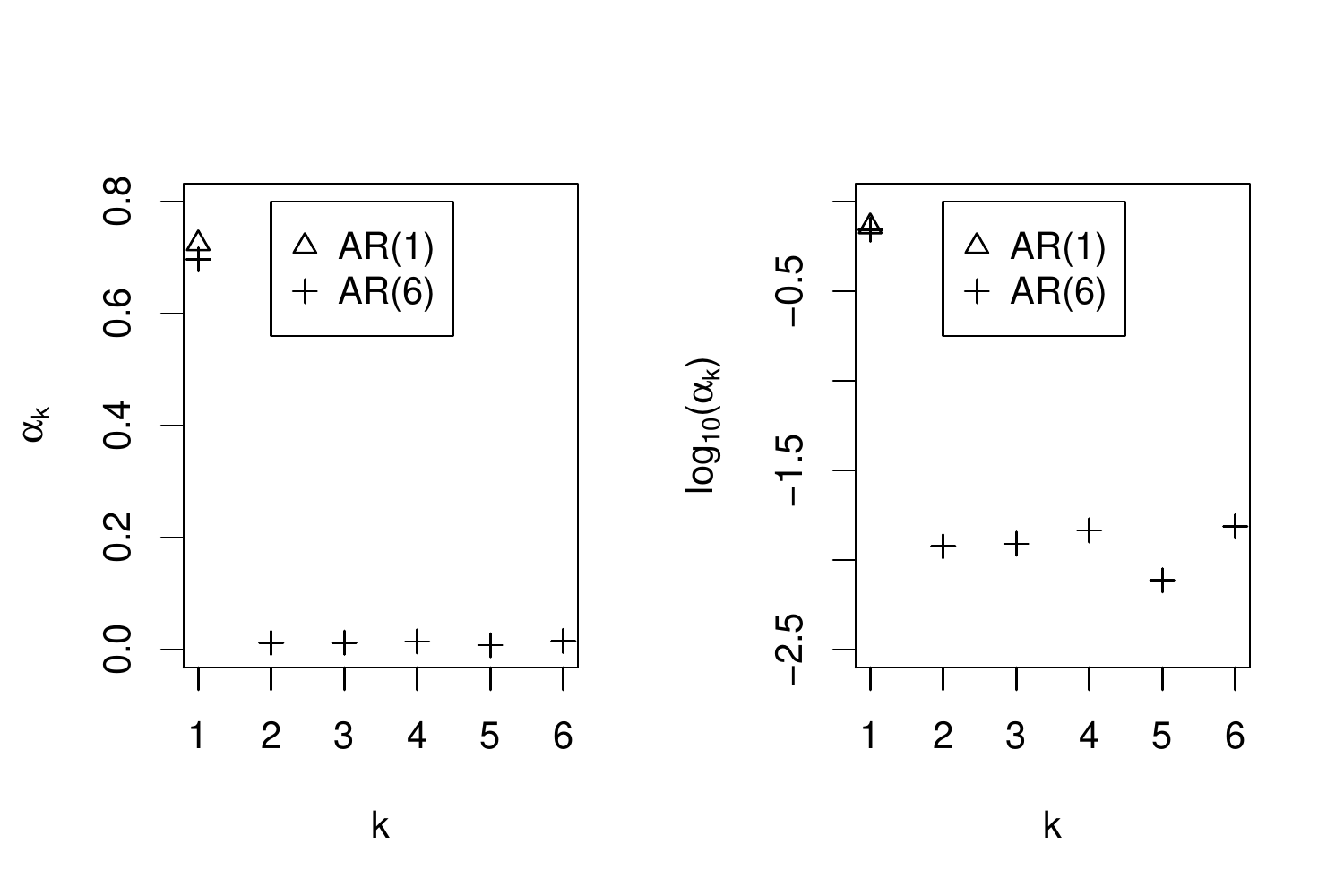}
\caption{Coefficients of autoregressive models of order one (triangle) and of
order 6 (crosses), plotted in normal (left) and logarithmic (right) ordinates.
In the AR(6) model, which is suggested as the optimal model by AIC, the
parameters $\alpha_2$ through $\alpha_6$ are on the order of $10^{-2}$ while
the parameter $\alpha_1$ is very close to that of the AR(1) model.}
\label{fig04}
\end{figure}

In \reffig{fig04}, it is shown that the parameters $\alpha_2$ through
$\alpha_6$ of the optimal AR(6) are only of the order of 0.01, while the first
parameter $\alpha_1$ is almost identical to that of the AR(1) process. We use
this as a motivation to override Akaike's suggestion and choose the AR(1)
process as our best data-driven model of the temperature anomalies. That is,
we model the temperature anomalies $T_n$ by
\begin{equation}
T_{n+1} = \alpha T_n + \epsilon_n,
\label{eqn04}
\end{equation}
where $\alpha=0.72$ and $\epsilon_n$ is Gaussian white noise with variance
$\sigma^2=3.06^2$.

In an AR(1) process with zero mean, parameter $\alpha$ and variance of the
residuals $\sigma^2$, it is straightforward to show that the marginal
distribution has mean zero and variance $\sigma_C^2=\mathbb{E}(T^2)$ equal to
$\sigma^2/(1-\alpha^2)$. Using our parameter estimates of $\sigma$ and
$\alpha$, we get $\sigma_C^2 = 4.42^2$ which is in agreement with the variance
of the anomaly distribution shown in \reffig{fig02}.

From \refeq{eqn04} one can conclude that, in an AR(1) process, the probability
distribution of the state at time instance $n+1$, conditional on the state at
instance $n$ is a Gaussian with mean equal to $\alpha T_n$ and variance equal
to $\sigma^2$, that is
\begin{equation}
(T_{n+1} \mid T_n) \sim \mathcal{N}(\alpha T_n,\sigma^2).\label{eqn05}
\end{equation}
We denote the Gaussian probability distribution function and cumulative distribution function by 
\begin{equation}
\varphi_{\mu,\sigma}(x)\equiv\frac{1}{\sqrt{2\pi}\sigma}\exp\left(-\frac{(x-\mu)^2}{2\sigma^2}\right),
\end{equation}
and
\begin{equation}
\Phi_{\mu,\sigma}(x)\equiv\int_{-\infty}^xdt\ \varphi_{\mu,\sigma}(t),
\end{equation}
respectively. If subscripts are missing, the conventions $\Phi\equiv\Phi_{0,1}$
and $\varphi\equiv\varphi_{0,1}$ apply. According to \refeq{eqn05}, the
probability of exceeding a threshold $\tau$ in an AR(1) process, conditional on
the present value $T_n$, is given by 
\begin{equation}
\mathbb{P}(T_{n+1} > \tau \mid T_n=t) = 1-\Phi_{\alpha t,\sigma}(\tau)\label{eqn07}
\end{equation}
If the true process that generates $T_{n+1}$ is indeed an AR(1) process,
\refeq{eqn07} provides the most complete information as to the occurrence of an
exceedance event.

\subsection{The weather model}\label{sec:raw}

The physical processes in the atmosphere are pretty well understood, although
not in full detail (e.~g. \cite{ATMbooks}). General circulation models (GCMs) are
models based on the hydrodynamic transport equations for the wind field plus
the thermodynamics of the transported air masses and their interaction through
the temperature dependent density of air. For more realism, further processes
have to be included, such as transport of different phases of atmospheric water
and their transitions, the energy budget has to be adjusted, topography must be
included, just to mention some. For detailed descriptions of state-of-the-art
atmospheric models see \cite{ecmwf2009}, \cite{noaa2011}, or \cite{dwd2012}.

For the forecast of temperature anomaly exceedances we use output from the NCEP
reforecast project \citep{hamill2005reforecast}. The reforecast project
provides a dataset of global ensemble weather forecasts. In a long reforecast
project, global temperature forecasts were issued using the same computational
model for the period 1979-present. I.~e., although this model was truly
operational only for a few years, it has been employed {\sl a posteriori} to
perform predictions on past observations, and it has been continued to perform
predictions until today even if since long better models have been available.
This is an invaluable source of data, since serious statistical analysis of
forecasts is possible if the same model is operated for several decades.
Initialized daily at 0:00 UTC, the model outputs forecasts on a
$2.5^\circ\times 2.5^\circ$ grid in 12-hourly intervals up to 15 days into the
future. An ensemble of 15 forecasts is produced by slightly varying the initial
conditions using so-called Bred perturbations \citep{toth1997ensemble}. See
\cite{leutbecher2008ensemble} for a review of methods and applications of
ensemble forecasting.

In order to issue temperature anomaly exceedance forecasts for Hannover, using
the ensemble forecast, we proceed as follows. Hannover's geographical coordinates
are 52.37N, 9.73E and the NCEP model has a grid point very close to these
coordinates, namely at 52.5N, 10.0E. We use the values of the ensemble
members at this grid point as an ensemble forecast for Hannover. We subtract
from the ensemble members the climatology in order to transform the
temperature forecast into an anomaly forecast. In the data-driven forecast,
we used today's measurement to estimate the probability of occurrence of an
exceedance event 24 hours in the future. Here, we use the 36 hours lead time
model forecast, in order to account for the time lag between measuring the
present state and actually having access to the model results.

\begin{figure}
\centering
\includegraphics[width=0.7\columnwidth]{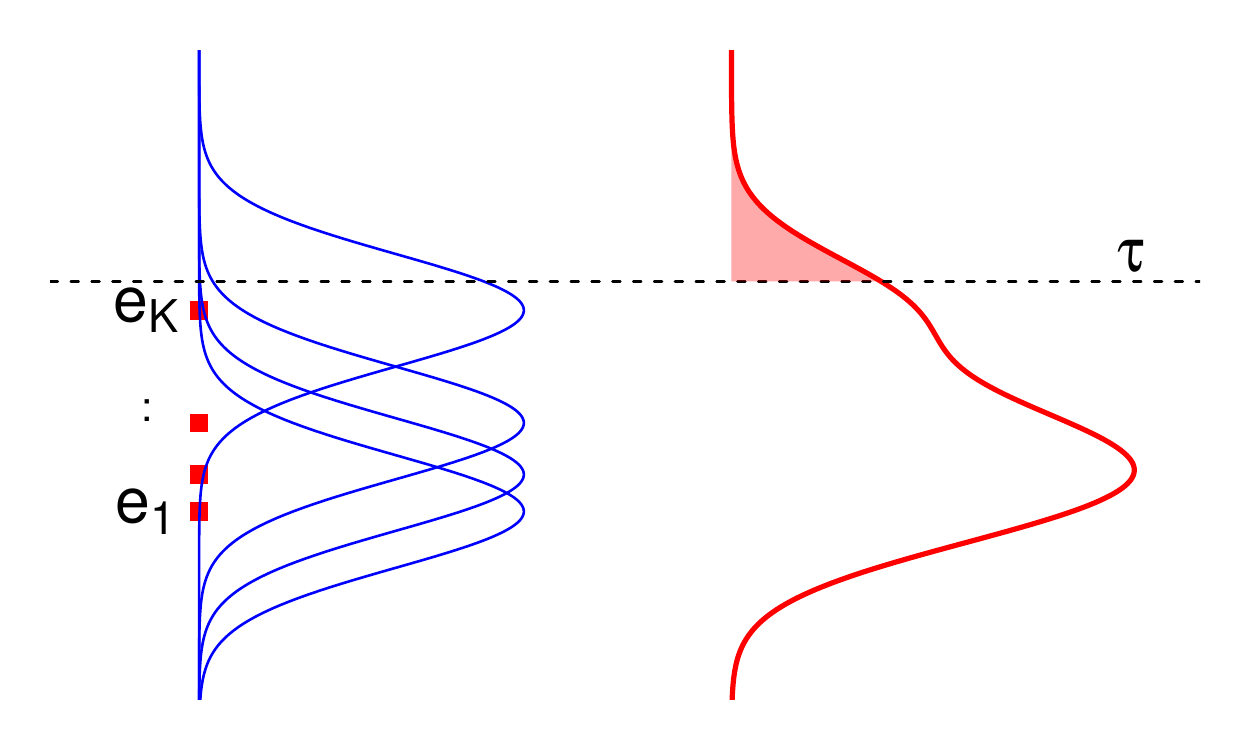}
\caption{Illustration of ensemble dressing. Each ensemble member (red markers) is dressed with a Gaussian kernel of zero mean and width $\sigma_k$ (blue lines). The superposition of all the dressing kernels provides the predictive distribution (red line). From this distribution, the exceedance probability of a threshold $\tau$ can be calculated (red shaded area).}
\label{fig09}
\end{figure}

In a first approach, we transform the ensemble into a predictive
distribution function by Gaussian kernel density estimation, the same method
that we used to estimate the anomaly distribution in \refsec{sec01}. That is,
we convert the discrete set of predicted temperature anomaly values into a
continuous probability density function. Applied to ensemble forecasts, kernel
density estimation is also referred to as ensemble dressing. Each ensemble member
is dressed with a Gaussian kernel function with zero mean and width $\sigma_k$,
which we calculate by Silverman's rule of thumb \citep{silverman1998density}.
For an ensemble of size $K$ and standard deviation $\tilde{\sigma}$, this rule
estimates the dressing kernel width $\sigma_k$ as
\begin{equation}
\sigma_k = \left(\frac{4\tilde{\sigma}^5}{3K}\right)^{\frac{1}{5}}.\label{eqn27b}
\end{equation}
Once the kernel width is estimated, the ensemble $\mathbf{e}=(e_1,\cdots,e_K)$ is
transformed into a density for the temperature anomaly by
\begin{equation}
p(T\mid \mathbf{e}) = \frac{1}{K}\sum_{i=1}^K \varphi_{e_i,\sigma_k}(T).
\end{equation}
\reffig{fig09} illustrates this method, as well as the calculation of the
exceedance probability of a threshold $\tau$. Note that, unlike suggested by
\reffig{fig09}, the ensemble members do not have to be ordered. We refer to the
above method of obtaining the exceedance probabilities as the raw ensemble
forecast.

\subsection{Calibrated weather model\label{sec:MOS}}

When using only the raw ensemble predictions, we ignore a very important point
concerning physical dynamical forecast models, namely that past prediction
errors can (and should) be used to improve future forecasts. With this insight
we enter the world of model output statistics (MOS;
\cite{glahn1972use}, \cite{wilks2007comparison}).

\begin{figure}
\centering
\includegraphics[width=0.7\columnwidth]{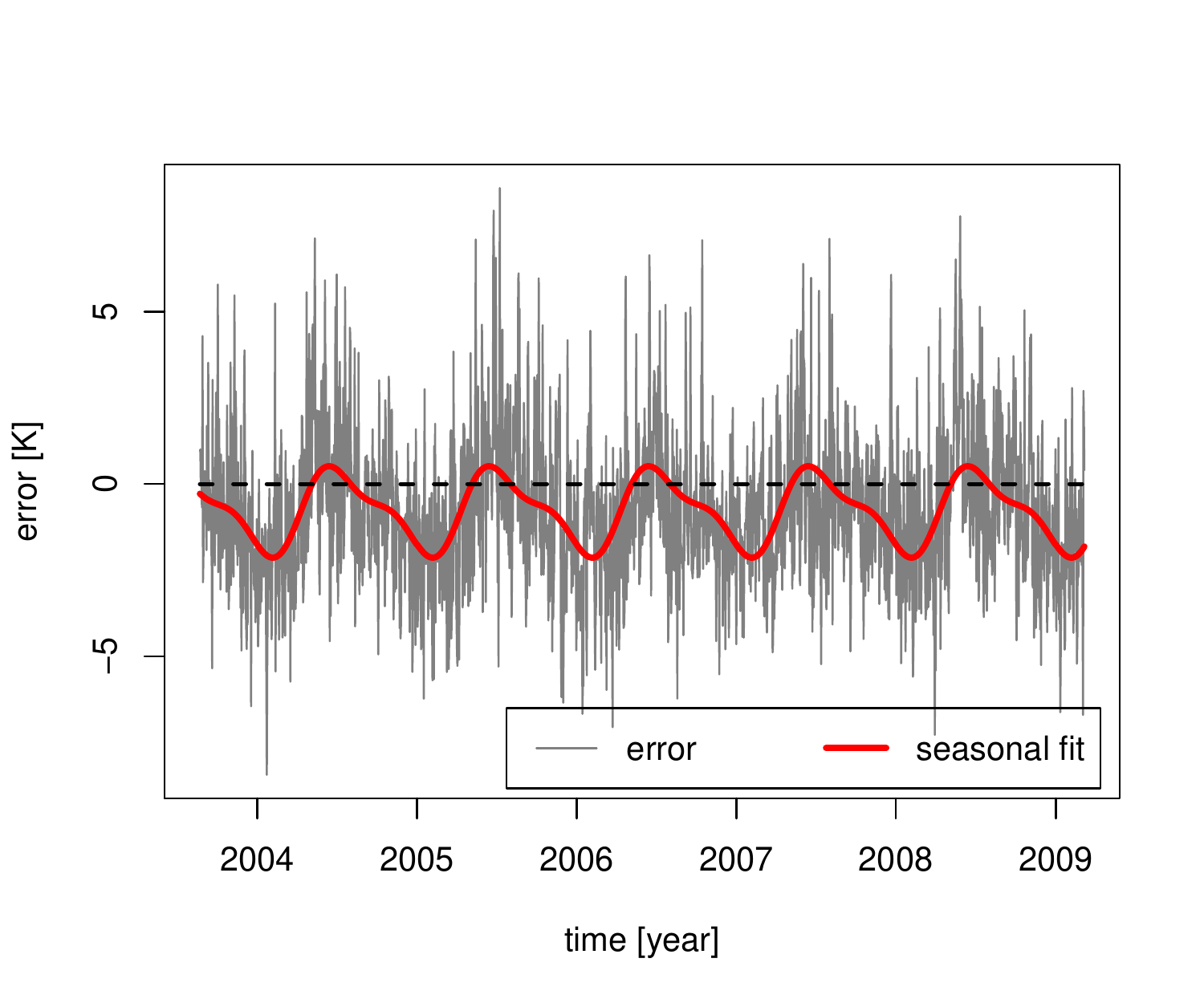}
\caption{\label{fig:bias}The difference between 
mean value of the NCEP ensemble forecasts and observation varies
systematically with season and its seasonal fit defines a bias which can 
be subtracted in order to improve forecast accuracy.
}
\end{figure}

The numerical model is only a sketch of the true atmosphere and thus model
errors are inevitable. However, some of these model errors are systematic, such
that they can be corrected for. Two notorious systematic errors in weather
models are seasonal bias and underdispersiveness. The bias is the average
difference between ensemble mean and verification, which is non-zero and
displays seasonality in the NCEP model (see \reffig{fig:bias}).
Underdispersiveness means that the ensemble variance underestimates the mean
squared difference between ensemble members and verification. Both model
errors are prevalent in the ensemble, and in the following we correct for
both of them. 

Here we employ one of different possible calibration schemes. It shifts the
values of every ensemble member by the same season-dependent value and corrects
the ensemble dispersion by an adjustment of the width of the dressing kernels.
More precisely, in analogy to our fitting of the climatology to the temperature
data in \refsec{sec01}, we fit a second order trigonometric polynomial to the
time series of the bias. The ensemble is bias-corrected by shifting the
ensemble mean according to the seasonal bias known from the two years preceding
the year of the forecast. 

In order to correct for ensemble underdispersiveness, we inflate the width of the
Gaussian kernels. 
In \cite{wang2005improvement} a method was proposed to estimate the kernel width for underdispersive ensembles under a second moment constraint.
Denote by $\overline{d^2}$ the average squared difference between ensemble mean and verification, by $\overline{s^2}$ the average ensemble variance, and by $K$ the number of ensemble members.
The kernel width proposed by \cite{wang2005improvement} is then given by
\begin{equation}
\sigma_k^2 = \overline{d^2} - \left(1+\frac{1}{K}\right)\overline{s^2}.
\end{equation}
With these model corrections which require and archive of past observations and
forecasts we can perform improved temperature anomaly exceedance forecasts.
Clearly, as in the AR(1)-model, we respect causality and we use only past data
for our re-calibration. The exceedance predictions are calculated as for the
raw ensemble, but after correcting for the bias and inflating the dressing
kernels. We refer to these predictions as the post-processed ensemble
forecasts.

\section{Probabilistic prediction of extreme anomalies}

All of our four models can be used to issue forecasts of the probability that
the temperature anomaly will exceed a predefined threshold on the next day.
Before we can compare the performances of these different models, we have to
define how to measure the skill of a probabilistic forecast.

\subsection{Scoring rules and the Brier Skill Score}\label{skillscores}
One way to evaluate probabilistic predictions is by means of strictly proper
scoring rules \citep{gneiting2007strictly}. A scoring rule is a function
$S(p,X)$ that combines a probabilistic forecast $p\in [0,1]$ and the
corresponding event indicator $X\in \{0,1\}$, where $X=1$ if the event happens
and $X=0$ otherwise. The scoring rule is proper if it forces the forecaster to
issue his probability honestly. Take the Brier Score
\citep{brier1950verification}, for example, which is given by 
\begin{equation}
\text{Br}(p,X)=(X-p)^2.
\end{equation}
The Brier Score is negatively oriented and zero for a perfect forecast that
assigns probability 1 to an event that actually occurs, probability 0 to an
event that does not occur. A forecaster who thinks that the probability of
occurrence of $X$ is $p$ can choose to issue a probability $q$ as his
forecast. He can calculate his subjective expectation value of the Brier Score
of the forecast $q$ by
\begin{equation}
\mathbb{E}(X-q)^2 = p(1-q)^2 + (1-p)q^2,
\end{equation}
where he assumes that the true rate of occurrence is his own estimate $p$. This
expectation is minimized if and only if $q=p$ which makes the Brier Score a
strictly proper scoring rule, i.~e., the forecaster has no chance to improve his
score by issuing a forecast $q$ that is different from his best guess $p$. The
same reasoning applies for the following scenario: Let $p$ be the true rate of
occurrence, and let $q$ be the best guess of the forecaster. Then that
forecaster performs best whose estimate is closest to the true value. Let us
stress that there are other, at first sight equivalent scoring rules, which
lack this property: replacing, e.~g., $(X-q)^2$ by $|X-q|$ leads to an improper
score, which can be improved by predicting $q=1$ whenever $p>1/2$ and $q=0$
otherwise. Propriety of a scoring rule is thus a reasonable property to ask
for. A further popular example of a strictly proper scoring rule is the
Ignorance Score \citep{roulston2002evaluating}, given by $-\log_2(p(X))$.

In the following we will compare different probabilistic forecasting schemes by
means of the Brier Score. A common way to compare scores of different
forecasts is by means of a skill score \citep{wilks2006statistical}. Let
$\bar{S}_1$ and $\bar{S}_2$ be the empirical averages of the Brier Score of
forecasting schemes 1 and 2, respectively. Then the Brier Skill Score (BSS) is
defined by
\begin{equation}
\text{BSS} = 1-\frac{\bar{S}_1}{\bar{S}_2}.\label{eqn18}
\end{equation}
The Brier Skill Score indicates the fraction of improvement of forecasting
scheme 1 over scheme 2 in terms of the Brier Score. 
A BSS of one indicates that the forecasts issued by scheme 1 are perfect,
i.~e., the forecast probability is unity 
each time the event happens and is zero each time
the event does not happen. 
A BSS of zero indicates no improvement and a negative BSS indicates that scheme 1 is inferior to scheme 2.

\subsection{Theoretical skill of the base rate model}

Before we test our models on the observed temperature anomalies, we compute the
theoretical values of the performance measures for our data models, i.~e., the
base rate model and the AR(1)-model. Let us assume for a moment that the temperature anomalies are really generated by an AR(1) process with parameters $\alpha$ and $\sigma$. We define the exceedance threshold $\tau$ to be the $q$-quantile of the climatological distribution of the temperature anomalies. As argued before, this distribution has zero mean and variance $\sigma_C^2=\sigma^2/(1-\alpha^2)$. Thus $\tau$ is defined such that 
\begin{equation}
q=\Phi_{0,\sigma_C}(\tau)=\Phi\left(\frac{\tau}{\sigma_C}\right).\label{eq:pquant}
\end{equation}

Averaged over all observations, a fraction of $(1-q)$ of the temperature
anomalies will be larger than $\tau$. However, we only issue predictions on
days when the temperature anomaly is below $\tau$. Under this constraint, the
average event rate $r_\tau$ is not equal to $(1-q)$, as the following
calculations show. The event rate $r_\tau$ in our setting is given by
\begin{equation}
r_\tau = \mathbb{P}(T_{n+1} > \tau \mid T_n \le \tau)
\end{equation}
which can be estimated from a data set using \refeq{eq:empiricalrate}.
In a true AR(1) process, we can calculate $r_\tau$ as a function of the AR parameters as follows:
\begin{align}
r_\tau& = \mathbb{E}[\mathbb{P}(T_{n+1}>\tau\mid T_n=t)\mid t \le \tau]\\
&=\mathbb{E}\left[1-\Phi_{\alpha t,\sigma}(\tau) \mid t\le\tau\right]\\
&=1-\left(\Phi_{0,\sigma_C}(\tau)\right)^{-1}\int_{-\infty}^\tau dt\ \Phi_{\alpha t,\sigma}(\tau)\varphi_{0,\sigma_C}(t),\label{eqn16}
\end{align}
where we made use of \refeq{eqn07} and the fact that $T_n$ is marginally distributed according to the
climatological Gaussian distribution with zero mean and variance $\sigma_C^2$.
Note that \refeq{eqn16} is equal to $(1-q)$ only if $\alpha=0$. The probability $r_\tau$ provides the CEBR forecast in this setting.

\begin{figure}
\centering
\includegraphics[width=0.7\columnwidth]{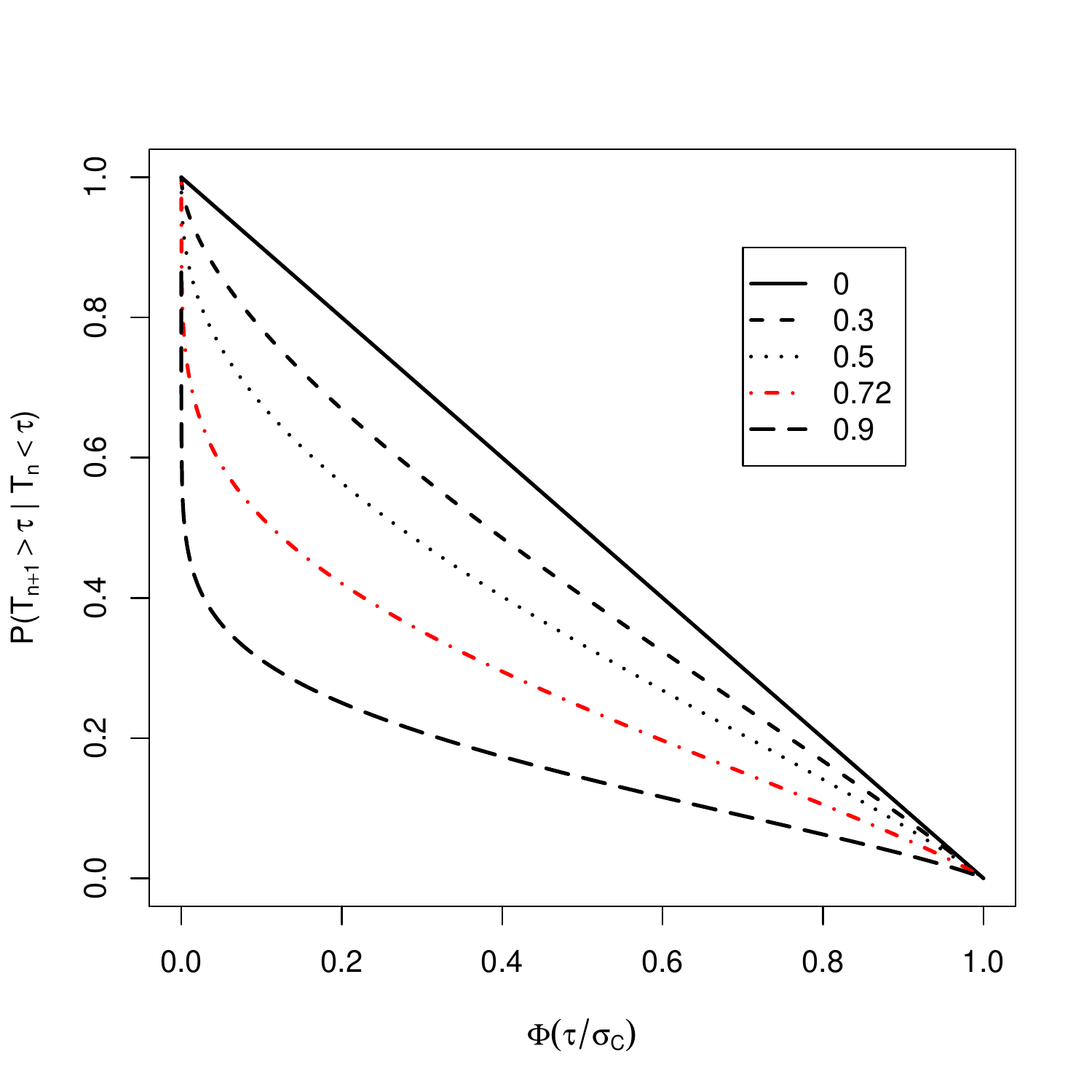}
\caption{Conditional exceedance base rate $r_\tau$ for AR(1)-data of a
threshold $\tau$ at instance $n+1$, conditional on not exceeding $\tau$ at
instance $n$, plotted over the $q$-value of $\tau$ in the climatological
distribution. Different lines indicate different values of the AR parameter
$\alpha$. The line that corresponds to our temperature anomaly time series is
shown red. Note that the climatological distribution is different for different
values of $\alpha$. As we define the threshold relative to the standard
deviation of the climatological distribution, these curves are independent
of $\sigma$.}
\label{fig05}
\end{figure}

We numerically integrated the expressions in \refeq{eqn16} (using the
\texttt{R}-function \texttt{integrate} provided by the \texttt{stats}-package,
\citep{Rmanual}) to produce the conditional exceedance rates in \reffig{fig05}.
The threshold $\tau$ that defines the exceedance event is defined with respect
to the climatological distribution which itself depends on the parameters of
the process $\sigma$ and $\alpha$. Its $q$-value is shown on the abscissa.
Since in all functions in \refeq{eqn16} the arguments are scaled by $\sigma$,
the curves of \reffig{fig05} do not depend on $\sigma$.

\reffig{fig05} shows that it might not be a good idea to issue $1-q$ as an
exceedance forecast if $\tau$ is the climatological $q$-quantile. Due to the
correlation of the process, the probability of hopping over the threshold,
conditional on being below the threshold at forecast time is reduced compared
to this probability in the uncorrelated process where $\alpha=0$. The process
has a tendency to stay below the threshold if it is already below the
threshold. This tendency is more pronounced, the higher the value of $\alpha$,
that is, the stronger the correlation. Since forecasts are only issued if the
present state is below the threshold, forecasting $1-q$ would overestimate the
CEBR if $\alpha > 0$.

The event $X:(T_{n+1}>\tau\mid T_n\le\tau)$ occurs with a rate $r_\tau$, given
by \refeq{eqn16}. The expectation value of the Brier Score of a probabilistic
forecast that constantly issues $r_\tau$ as a probability for $X$ is readily
calculated as follows:
\begin{align}
\mathbb{E}\text{Br}(r_\tau, X)&=(1-r_\tau)^2\mathbb{P}(X=1) + (0-r_\tau)^2\mathbb{P}(X=0)\\
&=r_\tau(1-r_\tau).\label{eq:CEBRbs}
\end{align}
This is the expected Brier Score of the CEBR forecast where the time series is
assumed to possess AR(1)-correlations and where the conditional base rate is
correspondingly smaller than one minus the probability corresponding to the
quantile. We will compare all further forecasts to this benchmark in terms of
the Brier Skill Score.

The expected Brier Scores given by \refeq{eq:CEBRbs} are shown as gray lines in
\reffig{fig06}. The maxima of all these curves assume the value 1/4, located
at that quantile where the conditional rate $r_\tau = 1/2$. 

We regard the CEBR as our null-model, the simplest possible prediction that a
forecaster who has access to a historical data set of temperature anomalies
could issue. A more sophisticated forecasting scheme would always have to be
compared to this simple null-model. We would only accept a more complicated
forecasting scheme if it can beat the CEBR forecast.

\subsection{Theoretical skill of the AR(1) model}

One forecast that is definitely more sophisticated than the CEBR forecast can
be obtained by issuing the true exceedance probability of the AR process at
the present value of $T_n$, namely $1-\Phi_{\alpha T_n,\sigma}(\tau)$ as given
by \refeq{eqn07}. As was mentioned before, this is the most complete
information as to the occurrence of an exceedance event in a true AR(1)
process. The expected Brier Score of this exceedance forecast is given by
\begin{align}
&\mathbb{E}\left[\text{Br}\left(1-\Phi_{\alpha T_n,\sigma}(\tau),X_{n+1}\right)\ \middle|\ T_n\le\tau \right]\nonumber\\
&=\left[\Phi_{0,\sigma_C}(\tau)\right]^{-1}\int_{-\infty}^\tau dt\ \varphi_{0,\sigma_C}(t)\left\{\Phi_{\alpha t,\sigma}(\tau)\left[1-\Phi_{\alpha t,\sigma}(\tau)\right]\right\}.\label{eqn27}
\end{align}
The term in the curly brackets of \refeq{eqn27} is the expected Brier Score at
a fixed value of $T_n=t$ and this term is averaged over all values of
$t\le\tau$, weighted by the marginal distribution. 

\begin{figure}
\centering
\includegraphics[width=0.7\columnwidth]{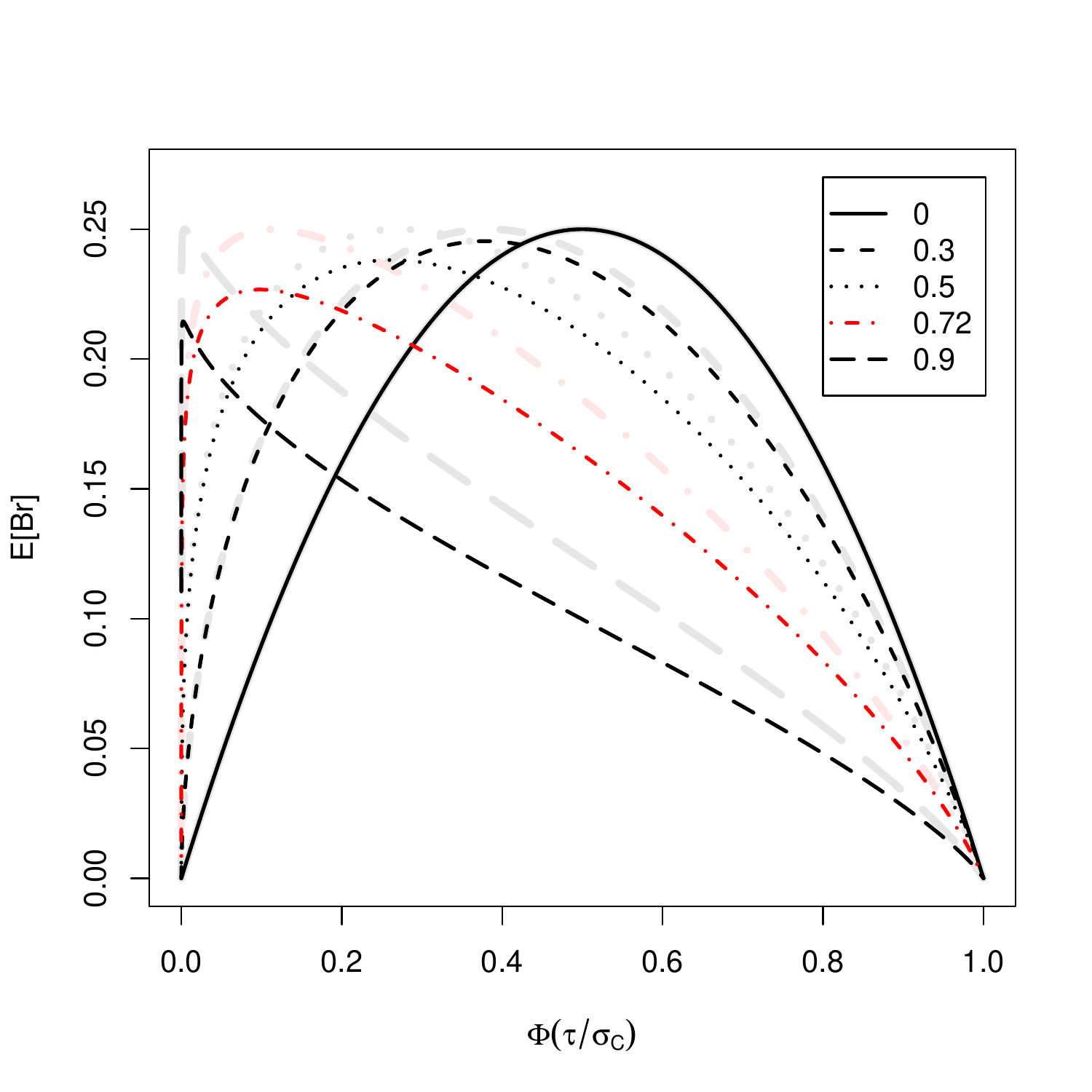}
\caption{Expected Brier Scores of the CEBR forecast (gray thick lines) and the true conditional exceedance forecast (black and red lines) for AR(1) processes with different AR parameters $\alpha$, as given by the legend.}
\label{fig06}
\end{figure}

We numerically integrate \refeq{eqn27} for different AR(1) parameters to
produce \reffig{fig06}. In an uncorrelated process, where $\alpha=0$, the
expected Brier Score of the true exceedance probability and that of the CEBR
are identical, because the two forecast probabilities are identical. If the
true process is uncorrelated, no prediction skill can be gained by assuming
correlation. For processes with $\alpha>0$, however, the expected Brier Score
of the CEBR is always higher (i.~e. worse) than that of the true exceedance
probability. Explicitly conditioning the forecast probability on the current
state $T_n$ leads to a significant gain in forecast skill. This gain is
monotonically increasing in the AR parameter $\alpha$. The maxima of all
curves occur at those points where the corresponding CEBR curves in
\reffig{fig05} cross the horizontal line $p=0.5$. At this point the uncertainty
of the forecaster as to the occurrence or non-occurrence of an exceedance event
is maximal, thus leading to the Brier Score being maximized. As $\tau$
approaches $+\infty$ or $-\infty$, all Brier Scores go to zero, that is, all
forecasts become more and more perfect. This can be seen as a result of the
growing certainty about the occurrence or non-occurrence of an exceedance event
if the threshold becomes ever smaller or larger.

\begin{figure}
\centering
\includegraphics[width=0.7\columnwidth]{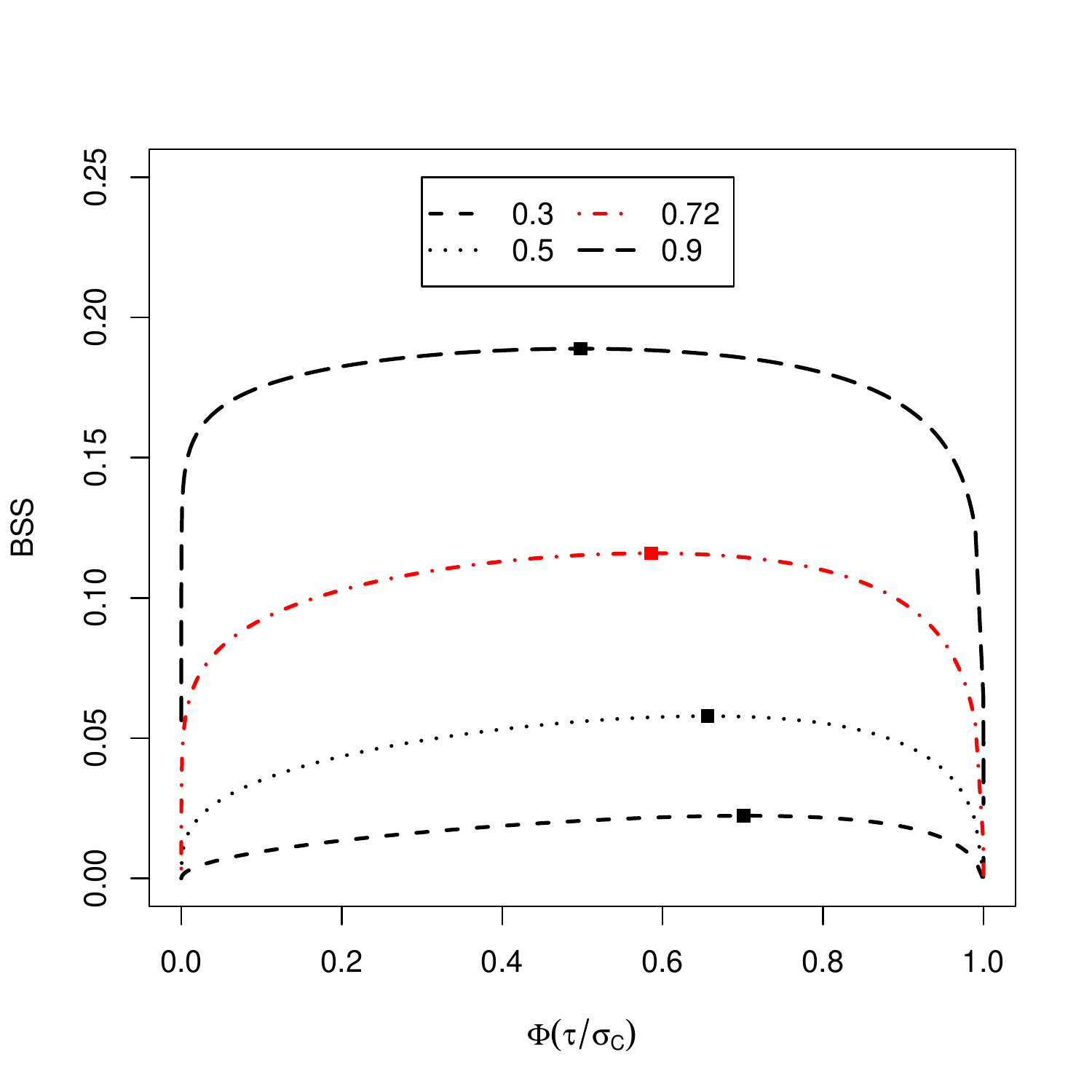}
\caption{Brier Skill Score comparing the forecasting scheme which forecasts the true exceedance probability to that forecasting only the CEBR for different AR(1) processes. The maxima are indicated by a marker.}
\label{fig07}
\end{figure}

Substituting \refeq{eq:CEBRbs} and \refeq{eqn27} into \refeq{eqn18} we compute
the Brier Skill Score that compares the Brier Score of the AR(1) forecast to
that of the constant CEBR forecast. Skill scores for different values of
$\alpha$ are shown in \reffig{fig07}. In the range between these extremes,
where $\tau$ is roughly between the 5- and the 95-percentile of the
climatological distribution of the process, the BSS is approximately constant.
In this range, the BSS is the larger, the larger the AR-parameter $\alpha$ is,
that is, the more correlated the process is.  While it is evident that the
Brier Score tends to zero when the event rate tends to either 1 or 0, it is
less evident that the Brier Skill Score for the AR-model does the same. The
curves shown in \reffig{fig07} are generated by numerical integration and seem
to converge to zero for large and small $\tau$ but we do not have any analytical
estimates for the Brier Skill Score in these limits. 

In Figs.~\ref{fig05}, \ref{fig06}, and \ref{fig07} the red lines report 
the theoretical results for that value of the AR-parameter $\alpha$ which we 
obtain by a fit of an AR(1) model to our temperature anomaly data. Hence, 
we expect the empirical skill of the AR-model on these data 
to be discussed in the next section to be similar.

\subsection{Empirical skill of the AR(1) model}
We now issue AR(1)-model predictions for the Hannover temperature
anomaly time series and compare these predictions to CEBR forecasts. We use
the time period between 1946-1978 (inclusive) to estimate the climatology, the
AR(1) parameters $\alpha$ and $\sigma$ as well as the CEBR as a function of
the threshold $\tau$. Based on this information we issue probabilistic
predictions for the event that a threshold $\tau$ will be exceeded by the
temperature anomaly, using the CEBR and using the AR(1) model. We compare the
probabilistic predictions to the actual outcomes of the events using the Brier
Score. Substituting the empirical averages of these scores into \refeq{eqn18}
yields the Brier Skill Score.

\begin{figure}
\centering
\includegraphics[width=0.7\columnwidth]{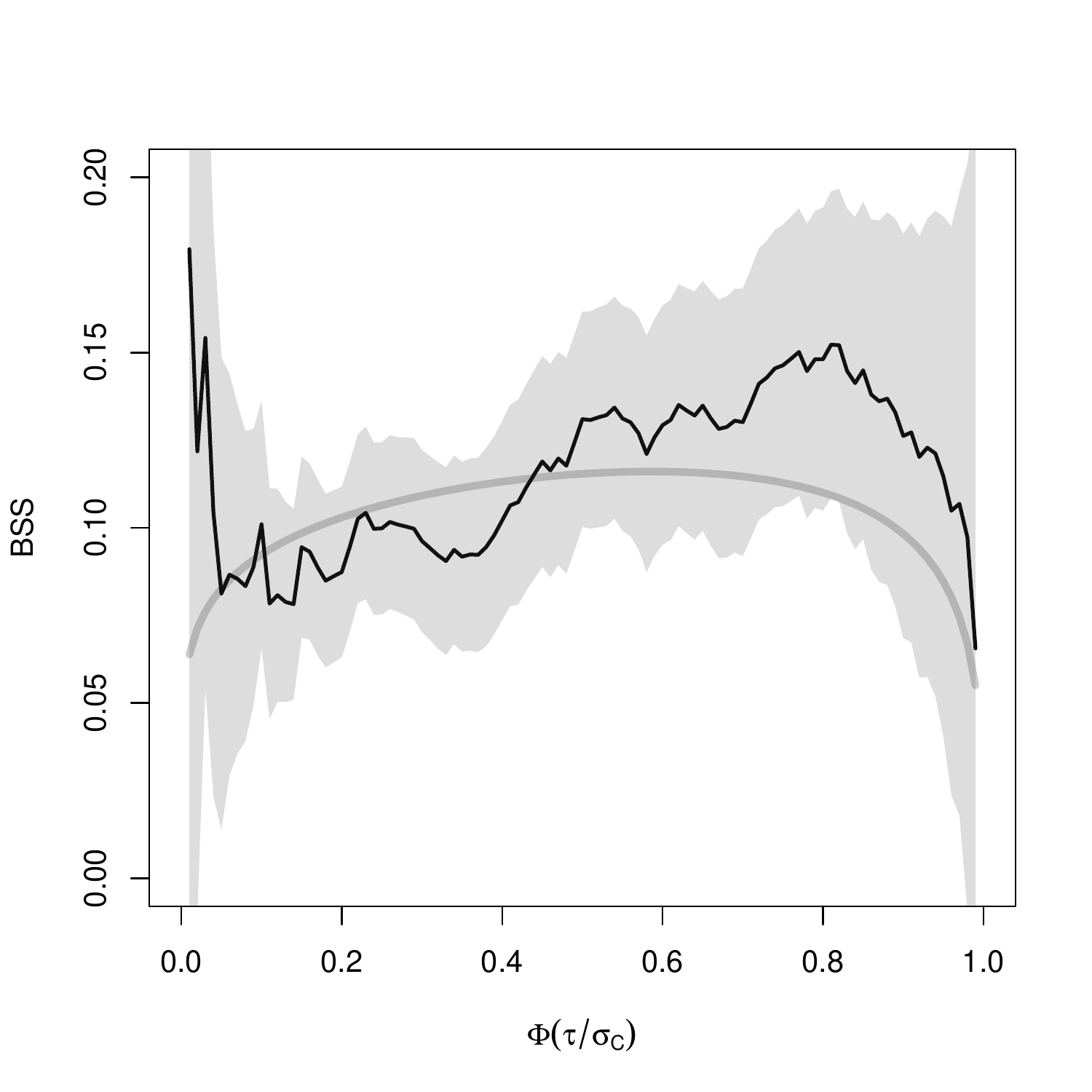}
\caption{Brier Skill Scores comparing the AR(1) forecast to the CEBR forecast in a true AR(1) process (grey line) and in the Hannover temperature anomaly time series (black line with 95 percent confidence interval).}
\label{fig08}
\end{figure}

This Brier Skill Score comparing the predictions issued by the AR(1) model to
those of the CEBR model is shown in \reffig{fig08} and compared to the
analytical result provided by comparing \refeq{eq:CEBRbs} and \refeq{eqn27} for
$\alpha=0.72$, which is the empirical value of the AR-parameter. Obviously, in
the temperature anomaly time series, predictions can be issued as to the
occurrence of an exceedance event that are significantly more skilful than the
CEBR. This result holds for a wide range of threshold values. Only for very
large negative and positive anomalies does the confidence band overlap zero so
that we can not assume significant improvement of the AR forecast over the CEBR
forecast. The analytical curve is fully contained in the confidence band, thus
reassuring that the calculations above are correct, and that the temperature
anomalies can indeed be modeled by an AR(1) process. 

Regarding confidence intervals, note that in \reffig{fig08}, as well as in all
other Figures, the intervals are to be taken as pointwise confidence intervals,
and not as confidence bands for the complete curve. If the confidence intervals
referred to the complete curve, they would be much wider. This distinction is
especially relevant if the points along the curve are not independent, which is
clearly the case if they refer to predictive skill with respect to different
threshold values. If predictive skill is particularly good at a threshold value
of, say, 0.8, it is reasonable to assume that predictive skill at threshold
value 0.81 is also good.

\subsection{Skill of the raw and post-processed ensemble forecast compared to
the AR(1)-model}

We evaluate the exceedance forecasts produced by the raw and post-processed
ensemble, which are documented in \refsec{sec:raw} and \refsec{sec:MOS}. Using
the Brier Skill Score we compare their Brier Score to the Brier Score of the
CEBR forecast. For reference and comparison, we include the Brier Skill Score
of the AR(1) forecast.

\begin{figure}
\centering
\includegraphics[width=0.7\columnwidth]{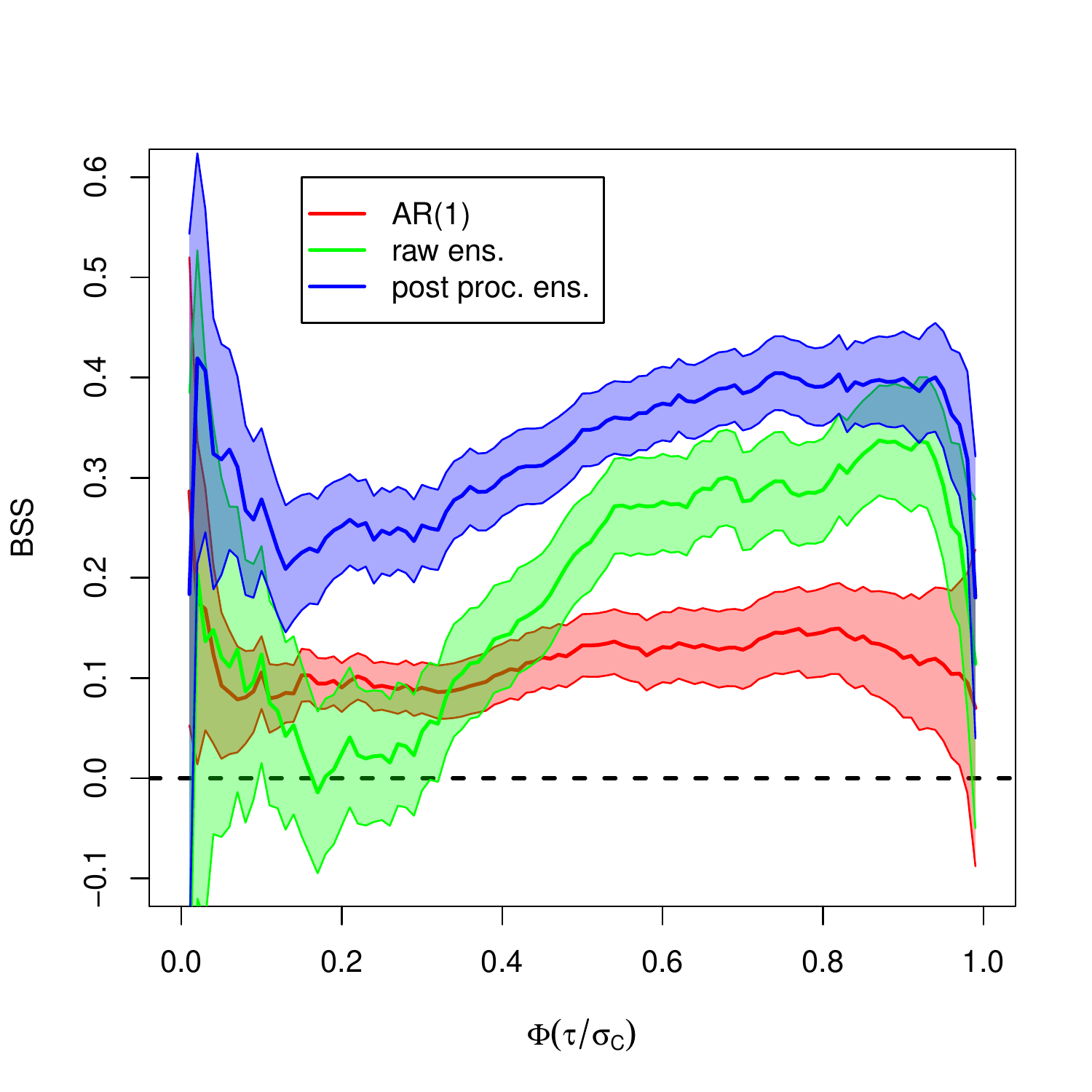}
\caption{Brier Skill Scores of the raw ensemble, the post-processed ensemble,
and the AR(1) forecast. 95 percent confidence intervals are included.}
\label{fig:BSS}
\end{figure}

\reffig{fig:BSS} shows, as a function of the threshold, the Brier Skill Scores of
the raw ensemble, of the post-processed ensemble, and of the
AR(1) forecast. For thresholds with $p$-values around 0.2, the ensemble can
hardly beat the CEBR, as indicated by a skill score close to zero. In this
range, the AR(1) forecast clearly outperforms the raw ensemble. For larger
thresholds on the other hand, the raw ensemble outperforms the AR(1)
predictions.

A valid question regarding \reffig{fig:BSS} is, how can it be that a complex
physical dynamical model is outperformed by the simple data-driven AR(1) model?
A weakness of the atmospheric model is that, unlike the AR(1) model, it is not
automatically calibrated to the observations. Systematic model errors, for
example due to unresolved topography, or errors in the estimation of the
initial model state can cause mis-calibration of the ensemble-based exceedance
forecasts, even though the weather model incorporates a thorough understanding
of the physical processes in the atmosphere. However, as discussed in
\refsec{sec:MOS}, observation data can be used to recalibrate the output of
the ensemble predictions.

We rerun the ensemble-based exceedance forecasts after applying bias correction
and variance inflation as described in \refsec{sec:MOS}. We model an
operational forecasting scenario by using data from the past two years to apply
corrections to forecast during a given year. That is, the seasonal bias as well
as the width of the dressing kernel are calculated using ensemble output and
observation data from the two years preceding the year of a given prediction.
By this procedure we account for non-stationarity in the model output, e.~g.
due to varying observation data. Note, however, that the post-processed
ensemble can only produce forecasts starting in 1981, since no calibration data
is available for the first two years in the data base.

\reffig{fig:BSS} shows Brier Skill Scores of the post-processed ensemble
forecast, taking the CEBR forecast as the reference forecast. The
post-processed ensemble forecast is constantly better than the AR(1) forecast,
as one would expect from a sophisticated physical dynamical model. The skill
score of the AR(1) forecast is exceeded by up to 0.3 by that of the
post-processed ensemble.  Furthermore, the ensemble post-processing
substantially improved the skill of the raw ensemble at all thresholds, most
remarkably at values of around $0.2$.

For very high and very low thresholds, the confidence intervals become very
wide. The skill scores of the three forecasts do not differ significantly. All
of them seem to tend to zero for very large thresholds. In this respect, the
ensemble forecasts share this property with the theoretical performance of the
AR(1) model shown in \reffig{fig07}.

The skill (or lack of skill) of probabilistic predictions can have different
causes. An additive decomposition of the Brier score was proposed by
\cite{murphy1973new} which quantifies two desirable forecast attributes, namely
reliability and resolution (also referred to as calibration and sharpness by
\cite{gneiting2007probabilistic}). We will not perform such an analysis of our forecasts here,
as it is the subject of a forthcoming paper. We note however, that the
forecasts produced by the raw ensemble are very unreliable, while the AR(1)
forecasts and the post-processed ensemble forecasts are almost perfectly
reliable for all values of the threshold $\tau$. The resolution of the two
ensemble forecasts is generally better than the resolution of the AR(1)
forecasts.

\section{From probabilistic to deterministic predictions of extreme anomalies} 

In certain situations, an end user might prefer a deterministic forecast over
a probabilistic one. This is in particular the case if the specific action
which the end user has to take in response to the forecast does not allow
for a gradual adjustment, but consists of exactly two alternatives.
Such a situation is typical of extreme weather: If, e.~g., a public event 
is sensitive to strong wind gusts, the two possible actions in response to 
the forecast ``probability $p$ for thunderstorm'' are only to ignore this 
danger or to cancel the event.

If such a decision has to be made repeatedly under a constant cost/loss
scenario, the end user will fix a certain threshold $\zeta$, and will act as if
the forecast was a deterministic ``yes'' if the predicted probability $p_n$ is
larger than $\zeta$. If $p_n < \zeta$, the end user will act as if a ``no'' was
predicted. A systematic way to evaluate such predictions for different values
of $\zeta$ is ROC analysis.

\subsection{The ROC analysis}\label{sec:ROC1}

ROC (Receiver Operating Characteristic) analysis \citep{Egan} is a performance
analysis for evaluating binary predictions (0/1), unfortunately without a
straightforward generalization to more than two classes. ROC analysis was
originally introduced in signal processing: Assume that a binary signal
(high/low) is sent over a noisy channel. The receiver has the task to
reconstruct the alternation of high/low by using an adjustable threshold. The
noisy channel leads to errors in this reconstruction.

Translated into prediction, we assume that the observations $X$ are either
``0'' or ``1'', and that the predictions $Y$ are as well either ``0'' or ``1''.
If predicted value and observed value coincide, $X_n=Y_n$, this prediction was
evidently successful. However, there are two different types of potential
mis-prediction: The forecast can be (a) $Y_n=1$ and the observation $X_n=0$ or
(b) vice versa. Skill scores such as the root mean squared prediction error
would weight these two errors identically. In many applications, and in
particular for extreme event prediction, but also in medical screening, this
can be very misleading: The real world costs for a missed hit (case b) are
usually very different from a false alarm (case a). Also, if the event rate is
very small, optimization of the root mean squared prediction error might lead
to assigning a better score to the trivial prediction which says ``0'' all of
the time (no false alarms, only a few missed hits) than to one which makes a
fair attempt to predict some ``1'' and thus suffers from both types of errors.

In view of these complications, a commonly used performance measure for such
binary prediction is the ROC curve. It assumes that the prediction scheme
possesses a sensitivity parameter $\zeta$, by which the relative number of $Y_n=1$
predictions can be controlled. The ROC curve is a plot of the hit rate versus
the false alarms rate parametrized by the sensitivity parameter $\zeta$ ranging from
insensitive (no ``1'' predicted, i.~e., no false alarms, no hits) to maximally
sensitive (always ``1'' predicted, i.~e., full record of hits, but also maximum
number of false alarms). Formally, the hit rate $H(\zeta)$ is the probability
of issuing alarms at sensitivity $\zeta$, given that the event actually occurs: 
\begin{align}
H(\zeta) = \mathbb{P}(p_n > \zeta \mid X_n = 1)= \mathbb{P}(Y_n = 1 \mid X_n = 1) \approx \frac{\sum_n X_n Y_n}{\sum_n X_n},
\end{align}
and the false alarm rate $F(\zeta)$ the probability of alarms given that no
event occurs:
\begin{equation}
F(\zeta) =\mathbb{P}(p_n > \zeta \mid X_n = 0) = \mathbb{P}(Y_n = 1 \mid X_n = 0) \approx \frac{\sum_n (1-X_n)Y_n}{\sum_n (1-X_n)}.
\end{equation}

This scoring scheme has a number of advantages with respect 
to others: \\
(a) a simple benchmark is a predictor which, at a given rate,
produces $Y_n=1$ irrespective of any information, so that the 
pairs $(X_n,Y_n)$ consist of two independent random variables. The ROC
curve of this trivial predictor is the diagonal. Hence, the
ROC curve of every nontrivial predictor has to be above the 
diagonal. \\
(b) As the reasoning in (a) shows, there is no explicit dependence 
of the ROC curve on the event rate, in contrast to, e.~g., the Brier score.
Therefore, ROC curves are suitable to compare predictive skill of different 
event classes, which also differ in their base rate.\\
(c) If costs for individual false alarms and for missed hits can 
be quantified and are known, then one can determine the working
point of a predictor, i.~e., the optimal sensitivity which minimizes 
the total costs.

A widely used summary index of a ROC curve is the area under the curve (AUC,
\cite{Egan}), defined as
\begin{equation} 
\text{AUC}=\int_{0}^{1}dF\ H(F).
\end{equation}
Since the trivial ROC curve is the diagonal, the trivial AUC value, which
should be exceeded by a nontrivial predictor, is equal to $0.5$. The perfect
value of AUC equals unity and indicates a predictor that differentiates between
events and non-events perfectly. We apply ROC analysis in the following section
to deterministic predictions of temperature anomaly exceedances.

Note that measures like the hit rate or AUC are fundamentally different from
proper scoring rules, as they cannot assign a value to a single
forecast-verification pair and can therefore not be written as averages over
individual pairs. This renders a quantitative comparison between ROC analysis
and proper scoring rules difficult.

\subsection{Comparison of the four models by ROC}\label{sec:ROC2}

We now want to predict temperature anomaly exceedance events by deterministic
predictions of the yes/no type. In other words, each day we want to predict
either ``yes, next day's anomaly will exceed the threshold $\tau$'' or ``no, it
will not''. Evidently, there are many ways how one can arrive at such
predictions. The most trivial and least useful one would be to simply toss a
coin, in other words to issue alarms randomly with a certain rate. However,
this is not as useless as it seems, since this provides a benchmark for every
serious prediction attempt: A predictor has to perform better than coin tossing
in order to be useful. At a given exceedance threshold, the base rate model of
\refsec{sec:CEBR} pre-defines the rate at which the coin should predict
``yes''. In the ROC analysis we can try all possible rates. But since all these
predictions are independent of the events, we create a diagonal line in the ROC
plot, according to the arguments in \refsec{sec:ROC1}. 

For the three other models, we convert the predicted probabilities into
deterministic yes/no predictions by the very simple rule mentioned above: Let
the predicted probability by either the AR(1) forecast, or the raw ensemble, or
the post-processed ensemble be $p_n$. We issue $Y_n=1$ if $p_n > \zeta$ and
$Y_n=0$ otherwise. The threshold $\zeta$ adjusts the sensitivity: If $\zeta$
close to 1, then very few $Y_n$ will be set to 1, whereas for $\zeta$ close to
zero $Y_n=1$ on many occasions. One might speculate that setting $\zeta$ such
that the relative number of 1's among the $Y_n$ is the same as among the
verifications $X_n$ is somehow optimal. Actually, in terms of calibration, this
would be the best choice, but in practice different values of $\zeta$ might be
preferred, as we will discuss below. In the following we use the algorithm
presented in \cite{fawcett2006introduction} to calculate ROC curves. AUC's and
their confidence intervals are calculated according to \cite{delong1988cta}.

\begin{figure}
\centerline{\includegraphics[width=0.7\columnwidth]{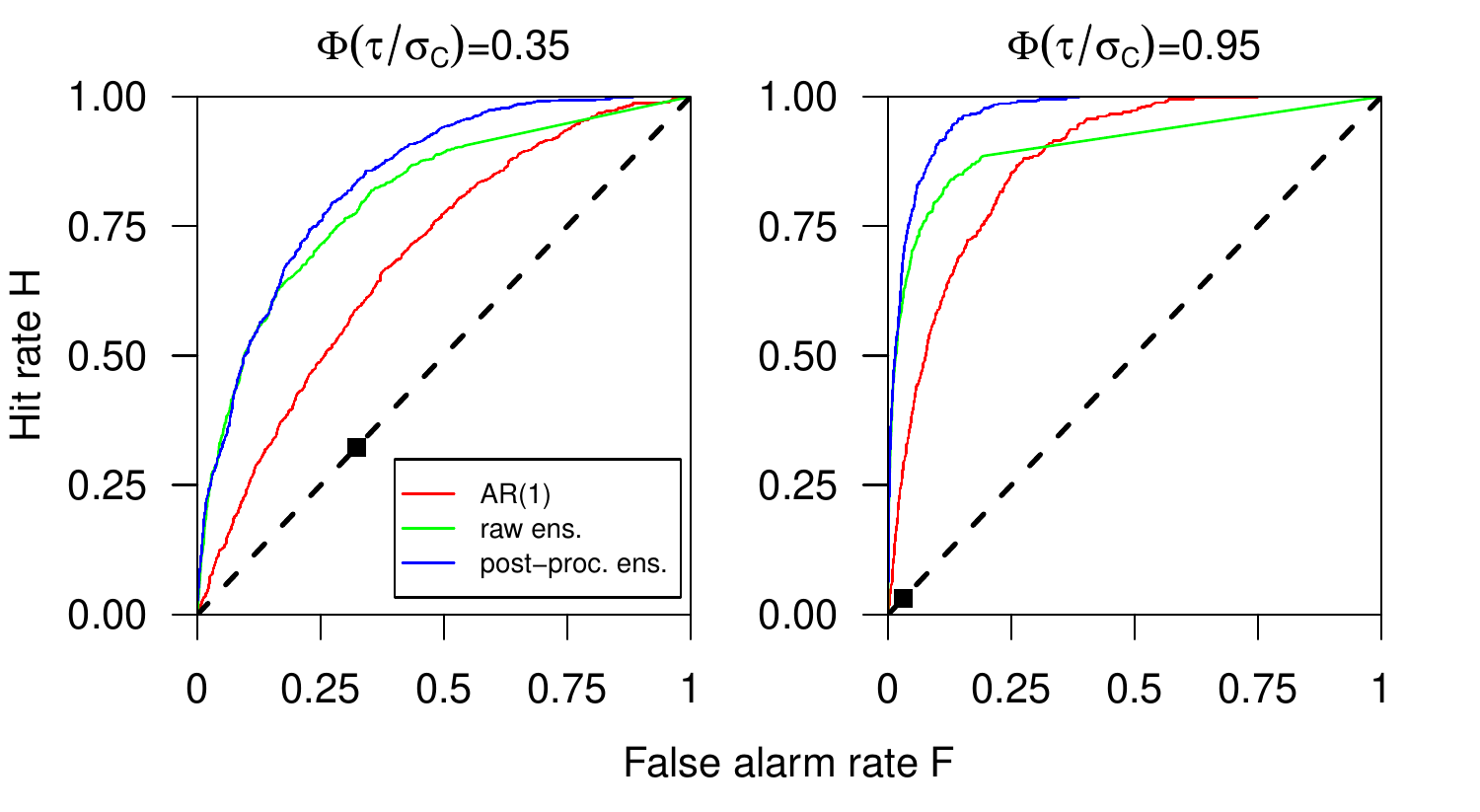}}
\caption{\label{fig:rocs}
ROC curves for deterministic predictions of temperature anomaly exceedances,
obtained by converting the probabilistic predictions issued by the AR(1) model
and the two versions of the ensemble forecast. Two exceedance thresholds are
shown: the 35-percentile, where the AR(1) model is superior to the raw ensemble
forecast in terms of the Brier Skill Score (\reffig{fig:BSS}) and the
95-percentile, which corresponds to exceedance events of very large thresholds.
The black squares denote the hit rate and false alarm rate of a predictor which
randomly issues alarms with a rate equal to the corresponding CEBR.
}
\end{figure}

In \reffig{fig:rocs} we show two ROC curves for deterministic exceedance
forecasts calculated from the probabilistic ones. For the first selected
$\tau$-value, the AR(1) model and the raw ensemble forecasts have about the
same predictive skill in terms of Brier Skill Score (\reffig{fig:BSS}).
Conversely, in terms of its ROC curve, the raw ensemble is closer to the
post-processed ensemble than to the AR(1) forecast. Once again, the
post-processed ensemble is superior to both. The second $\tau$-value
corresponds to exceedance events of very large thresholds. The ROC curves of
the three nontrivial prediction schemes are closer to the optimal point $(0,1)$
than at the smaller exceedance threshold $\tau$. However, at this larger
threshold and at low values of the sensitivity parameter $\zeta$, the ROC of
the AR(1) forecast lies above that of the raw ensemble forecast. At high false
alarm rates, the AR(1) based forecast has a higher hit rate than the raw
ensemble forecast.

The ROC plots in \reffig{fig:rocs} are typical of all other $\tau$-values, the
main variation being how closely the individual curves approach the desired
upper left corner $(H=1,F=0)$. A coin-tossing model, generating
$Y=1$-predictions with any rate, will generate the diagonal. The black squares
indicate the performance of such a model if the rate is taken as the true
conditional exceedance base rate, \refeq{eq:empiricalrate}. One possible
conclusion of this plot is: The base rate model causes a given percentage of
false alarms. If we accept the same number of false alarms for the more
sophisticated models, we have a much better hit rate. Or vice versa, the base
rate model has a given fraction of hits. If our improved models are to be
operated such that their hit rate is the same, then they would produce much
less false alarms.

\begin{figure}
\centering
\includegraphics[width=0.7\columnwidth]{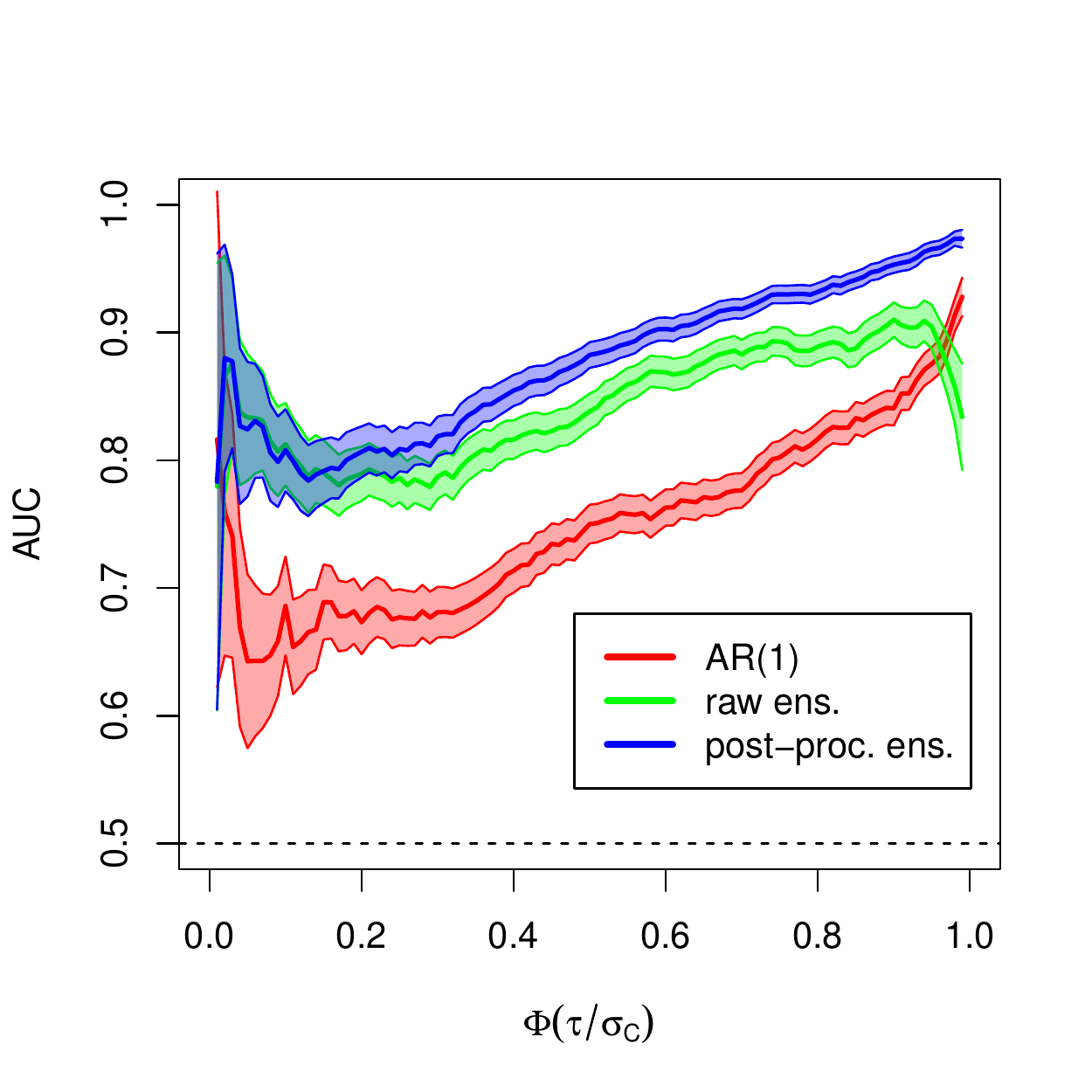}
\caption{\label{fig:AUC}The area under curve, AUC, as a function of the
exceedance threshold value, for the three models. 95 percent confidence
intervals are included. The more the AUC-value exceeds 1/2, and the closer it is
to the maximum of 1, the better the average performance.}
\end{figure}

In \reffig{fig:AUC} we report the dependence of AUC on the threshold $\tau$ for
the three nontrivial predictors. As for the Brier Skill Score
(\reffig{fig:BSS}) we observe systematic differences for different exceedance
thresholds $\tau$ and significant differences between the three prediction
schemes. There are, however, a number of notable differences to the BSS. At
the low thresholds, where the AR(1) model outperforms the raw ensemble in terms
of the BSS, the raw ensemble is much better in terms of the AUC, and even close
to the post-processed ensemble. At very large thresholds, on the other side,
the AR(1) model has a significantly higher AUC than the raw ensemble. This
deficiency is eliminated by the post-processing. Another interesting behavior
is the apparent increase of the AUC with increasing values of $\tau$. That
means that, the more ``extreme'' the events get, the better they become
predictable in the ROC sense. This effect has been previously observed for
different prediction targets in \cite{Hallerberg}. The BSS, on the other hand,
tends to zero for very large thresholds.

A possible explanation for these systematic differences between the evaluation
criteria BSS and AUC is as follows: We have introduced an additional parameter,
the sensitivity $\zeta$, which is a kind of implicit re-calibration. Assume a
probabilistic model and a modification of that by simply dividing all predicted
probabilities by two. The modified model would have exactly the same ROC curve
and AUC as the original one because these measures are invariant with respect
to monotonic transformations.  On the other hand, the modified model would
usually have a worse Brier Skill Score because the Brier Score is indeed
sensitive to such a transformation. One could therefore argue that ROC analysis
only measures forecast resolution, which is the higher, the better informed a
forecaster is. The increase of performance of the weather model becomes thus
more obvious. However, a formal connection between the AUC and forecast
resolution in the sense of the reliability-resolution-uncertainty decomposition
of the Brier Score \citep{murphy1973new} has yet to be established. 

The fact that the ensemble post-processing improves the ROC measures shows that
our ensemble post-processing is not the same as a simple linear recalibration
of the forecast probabilities. As stated above, ROC measures are invariant
under such a recalibration. But since we modify the raw ensemble and not the
forecast probabilities, and since the seasonal bias correction alters the
probabilities nonlinearly, we are able to significantly improve ROC curves and
AUC of the exceedance forecasts by the ensemble post-processing.

\section{Discussion and conclusions} 

We gave an overview over different forecast products related to extreme events.
The forecast itself, regardless of the specific forecast product, is an
input-output relationship. Depending on availability, one may use physical
dynamical models, statistical learning algorithms, or data based prediction
schemes in order to make use of input data. The evaluation of such predictions,
and hence the decision which forecast product is optimal for a given problem,
requires the definition of a performance measure. Since there are many
different possibilities for scoring, there might be several optimal predictors.

As a specific example, we discuss the prediction of temperature anomalies
exceedance events. We compare two prediction schemes: one results from a
dynamical weather model, the other is a simple time series model fitted to past
data of the measurement station under consideration. The difference in model
complexity, in computational effort, and in the dimensionality of input data is
tremendous. Nonetheless, the performance of the time series model is not as bad
as one might naively expect: The improvement over a benchmark predictor is in
some sense of the same order of magnitude. Interestingly, there are even
prediction tasks where the uncalibrated weather model performs worse than the
time series model.

In weather forecasting, the calibration of a dynamical model to the local
statistics is essential in order to provide good forecasts, because model
errors introduce systematic biases. In situations where such calibration
functions are unknown, such predictions may be systematically wrong and hence
misleading. Such lack of calibration is evidently given in areas of the world
where there are no measurement stations which might be used to calibrate the
local forecasts.  In view of climate change and extreme weather, this
calibration issue leads to another problem: How the model post-processing has
to be modified under changed climatic conditions can only be guessed.
Unfortunately, for the estimate of the relative frequency of extreme anomalies,
this calibration is essential, as can be seen in \reffig{fig:BSS}. Since
observation data for a climate different from the present one is unavailable,
the probability of extreme weather events can only be estimated by dynamical
models, even if they are not perfect.  In settings where a dynamical model is
unavailable due to lack of equations that describe the physics of the system,
data based modeling is a serious alternative; its performance might be better
than the performance of an uncalibrated dynamical model.

By converting predicted probabilities into deterministic binary forecasts
and evaluating these by ROC statistics, different properties of the forecast
scheme are evaluated. A violation of calibration becomes irrelevant, hence the
raw ensemble performs almost as well as the post-processed ensemble. 
The last item confirms what we said in the introduction: We can only
speak about the optimal predictor after we have decided how we wish to
evaluate the skill of predictions.

Extreme event prediction is rare event prediction, i.~e., prediction in
the limit of the base rate tending to zero. Hence, we should compare the Brier
Skill Score and the Area Under the Curve in the rightmost part of
Figs.~\ref{fig:BSS} and \ref{fig:AUC}, which both compare a more sophisticated
forecast to a base rate forecast. Whereas \reffig{fig:BSS} suggest that
predictability tends to disappear (no improvement over the trivial base rate
forecast), \reffig{fig:AUC} suggests the opposite: Events become the better
predictable the higher we adjust the threshold of what we call extreme.  This
contradiction shows how relevant the choice of the performance measure, and
related to that, the choice of the forecast product is.

The present study provides a number of directions for future studies: We have
only considered forecasts 24 hours ahead. We expect the scores of all
prediction models to decrease for higher lead times. We expect the raw ensemble
forecast to systematically outperform the AR(1) model at higher lead times.
Furthermore, the reasons for the low Brier Skill Score of the raw ensemble can
be worked out more carefully by a decomposition of the Brier Score into
reliability, resolution and uncertainty \citep{murphy1973new}. Lastly, we
compared rather simple forecasting models both from the data-based and from the
physical dynamical family of models. Neither the weather model nor the
data-driven prediction model can be considered state-of-the-art. The
performance of the NCEP reforecast model is surely not representative of
state-of-the-art weather models. The conclusions might change if different
prediction models are used.

\section*{Acknowledgments}

We would like to thank an anonymous reviewer for helpful comments.

\end{document}